\documentclass{article}

\usepackage{arxiv}

\usepackage[utf8]{inputenc} % allow utf-8 input
\usepackage[T1]{fontenc}    % use 8-bit T1 fonts
\usepackage{hyperref}       % hyperlinks
\usepackage{url}            % simple URL typesetting
\usepackage{booktabs}       % professional-quality tables
\usepackage{amsfonts}       % blackboard math symbols
\usepackage{nicefrac}       % compact symbols for 1/2, etc.
\usepackage{microtype}      % microtypography
\usepackage{lipsum}
\usepackage{graphicx}
\usepackage{amsmath}
\graphicspath{ {./images/} }

\title{Inspired by machine learning optimization: can gradient-based optimizers solve cycle skipping in full waveform inversion given sufficient iterations?}

\author{
 Xinru Mu \\
  Physical Science and Engineering Division\\
  King Abdullah University of Science and Technology\\
  Thuwal 23955, Saudi Arabia \\
  \texttt{xinru.mu@kaust.edu.sa} \\
   \And
 Omar M. Saad \\
  Physical Science and Engineering Division\\
  King Abdullah University of Science and Technology\\
  Thuwal 23955, Saudi Arabia \\
  \And
 Shaowen Wang \\
  Physical Science and Engineering Division\\
  King Abdullah University of Science and Technology\\
  Thuwal 23955, Saudi Arabia \\
  \And
 Tariq Alkhalifah \\
  Physical Science and Engineering Division\\
  King Abdullah University of Science and Technology\\
  Thuwal 23955, Saudi Arabia \\
}

\begin{document}
\maketitle

\begin{abstract}
Full waveform inversion (FWI) iteratively updates the velocity model by minimizing the difference between observed and simulated data. Due to the high computational cost and memory requirements associated with global optimization algorithms, FWI is typically implemented using local optimization methods. However, when the initial velocity model is inaccurate and low-frequency seismic data (e.g., below 3 Hz) are absent, the mismatch between simulated and observed data may exceed half a cycle, a phenomenon known as cycle skipping. In such cases, local optimization algorithms (e.g., gradient-based local optimizers) tend to converge to local minima, leading to inaccurate inversion results. In machine learning, neural network training is also an optimization problem prone to local minima. It often employs gradient-based optimizers with a relatively large learning rate (beyond the theoretical limits of local optimization that are usually determined numerically by a line search), which allows the optimization to behave like a quasi-global optimizer. Consequently, after training for several thousand iterations, we can obtain a neural network model with strong generative capability. In this study, we also employ gradient-based optimizers with a relatively large learning rate for FWI. Results from both synthetic and field data experiments show that FWI may initially converge to a local minimum; however, with sufficient additional iterations, the inversion can gradually approach the global minimum, slowly from shallow subsurface to deep, ultimately yielding an accurate velocity model. Furthermore, numerical examples indicate that, given sufficient iterations, reasonable velocity inversion results can still be achieved even when low-frequency data below 5 Hz are missing.
\end{abstract}


\section{Introduction}
Full waveform inversion (FWI) is a highly accurate method for estimating subsurface velocity models. It iteratively updates the velocity model to minimize the misfit between simulated and observed seismic data \cite{tarantola1984inversion}. However, when the initial velocity model is inaccurate, the phase difference between the observed and simulated data can exceed half a cycle, a phenomenon known as cycle skipping \cite{virieux2009overview}. In such cases, using a local optimizer for inversion is likely to lead to convergence to a local minimum, thereby reducing inversion accuracy. Because low-frequency data are less sensitive to cycle skipping, a multi-scale FWI strategy has been proposed, which performs inversion progressively from low-frequency to high-frequency data \cite{bunks1995multiscale}. This hierarchical approach helps to mitigate cycle skipping and facilitates the recovery of an accurate velocity model. Unfortunately, in practice, seismic data below 3 Hz are often unavailable due to limitations in instrumentation, environmental noise, and other factors \cite{chen2022cycle}. The absence of low-frequency signal hampers the reliable reconstruction of the low-wavenumber background model, and as a result, cycle skipping may still exist in subsequent high-frequency inversions. To further mitigate the problem of cycle skipping, many methods have been developed, including the use of geological and geophysical priors, advanced objective functions, global optimization methods, and machine learning–assisted inversion methods, among others \cite{asnaashari2013regularized, sen2013global, alkhalifah2016full, warner2016adaptive, yao2019tackling, li2021deep, chen2022cycle, zhao2022hybrid, sun2023implicit}.

Well logs provide accurate and reliable subsurface velocity information. Therefore, incorporating well log velocities as regularization terms in FWI can help prevent convergence to local minima \cite{asnaashari2013regularized}. Additionally, facies information extracted from well data can also be used as a regularization constraint to enhance the accuracy of FWI \cite{zhang2018multiparameter}. In recent years, deep neural networks (DNNs) have demonstrated strong capabilities in learning the distributions of well log velocities and facies, which are subsequently incorporated as regularization priors in FWI \cite{zhang2022regularized, li2021deep, wang2023prior}. 

The conventional FWI commonly employs the least-squares \({L_2}\)-norm for data misfit, which involves point-by-point comparison and is prone to cycle skipping. To overcome this limitation, some advanced objective functions have been proposed, many of which adopt trace-by-trace (or gather-by-gather) comparisons. Among these, the optimal transport objective function has shown strong potential in effectively alleviating the cycle skipping issue in FWI and has attracted significant research and practical interest \cite{engquist2016optimal, yang2018application, sun2019application, yong2019misfit, da2022graph}. It measures the difference between observed and simulated seismic data by computing the minimal cost required to transport one distribution into the other, thereby capturing both amplitude and phase discrepancies in a physically meaningful manner. Another promising approach is the matching filter-based misfit function, which computes a matching filter through deconvolution between observed and simulated seismic traces \cite{van2010correlation, luo2011deconvolution, warner2016adaptive, sun2019robust, yong2023localized}. The associated optimization problem is formulated by penalizing filter coefficients at non-zero time lags, as the ideal matching filter becomes a band-limited Dirac delta function when the simulated data perfectly aligns with the observed data. Furthermore, the dynamic time warping (DTW)-based loss function has proven effective in mitigating cycle skipping \cite{hale2013dynamic, yang2014using, chen2022cycle, tan2024full}. DTW works by nonlinearly aligning simulated and observed seismic traces in time, allowing for local stretching and squeezing to achieve optimal alignment between the signals. In recent developments, DNN–based objective functions have also been developed to improve the measurement of similarity between simulated and observed data, thereby enhancing the overall accuracy of FWI \cite{sun2022ml, yang2023fwigan, saad2024siamesefwi}.

Another way to solving cycle skipping is the use of global optimization methods, which explore the parameter space more comprehensively and are less likely to become trapped in local minima. Genetic algorithms, simulated annealing, particle swarm optimization, and Monte Carlo methods have all been successfully applied to FWI, yielding promising results \cite{sen1991nonlinear, sambridge1992genetic, jin1993background, sen2013global, aleardi2019assessing, mojica2019towards}. To improve the computational efficiency of global optimization methods, \cite{zhao2022hybrid} incorporated gradient information into the simulated annealing algorithm, resulting in a gradient-based global optimization approach that significantly reduces computational cost. Building on this idea, \cite{zhao2021gradient} further incorporated gradient information into the Markov Chain Monte Carlo (MCMC) method, resulting in a hybrid optimization approach that was successfully applied to both FWI and uncertainty analysis. More recently, \cite{berti2024probabilistic} employed gradient-based MCMC method for surface wave inversion and uncertainty quantification. However, due to the large number of iterations and high memory requirements, global optimization algorithms are rarely used in practical industrial applications. Taking into account this balance between global and gradient-based optimizers, a question may arise: If we have the luxury to push gradient-based FWI methods to global search cost levels and iterations, would we converge? 

In this paper, inspired by the optimization implementation in machine learning, where gradient-based optimizers combined with a relatively large learning rate can, after thousands of iterations (starting from a random initialization), produce a well-optimized neural network model. We apply this optimization strategy to FWI to investigate whether it can resolve cycle skipping. We conduct experiments on three velocity models: a linearly increasing model, the Overthrust model, and the Marmousi2 model, along with a field data test. Forward modeling is performed using the acoustic wave equation, with a free-surface boundary condition to simulate multiples. The synthetic shot gathers are filtered to remove frequency components below 3 Hz. For all FWI tests, we use a linearly increasing velocity model as the initial guess to intentionally introduce cycle skipping between the simulated and observed data. Numerical experiments show that, given a sufficient number of iterations, the gradient descent (GD) optimizer can overcome cycle skipping and yield accurate inversion results. To further accelerate convergence, incorporating momentum into the GD framework has proven effective by smoothing the optimization trajectory and helping escape local minima. Building on this, the Adaptive Moment Estimation (Adam) optimizer \cite{kingma2014adam}, which combines momentum with adaptive learning rates, further reduces the required number of iterations. Moreover, using the Adam optimizer with mini-batch shot gathers in each iteration further enhances efficiency by not only reducing the computational cost per iteration but also decreasing the number of iterations required to achieve accurate results.

\section{Theory}
FWI aims to reconstruct high-resolution subsurface velocity models by minimizing the misfit between observed and simulated seismic data. In this section, we briefly introduce the fundamental principles of acoustic FWI.

\subsection{Principle of acoustic FWI}
Given a 2D velocity model \(v\), we can consider that seismic wave propagation is governed by the acoustic wave equation, which can be written as:

\begin{equation}
\label{eq1}
\frac{{{\partial ^2}p(\textbf{r},t)}}{{\partial {t^2}}} = {v^2}(\textbf{r})\left( {\frac{{{\partial ^2}p(\textbf{r},t)}}{{\partial {x^2}}} + \frac{{{\partial ^2}p(\textbf{r},t)}}{{\partial {z^2}}}} \right) + f({\textbf{r}_s},t)\delta (\textbf{r} - {\textbf{r}_s}),
\end{equation}
where \({p(\textbf{r},t)}\) is the pressure wavefield at spatial coordinate vector \({\bf{r}} = \left( {x,{\rm{ }}z} \right)\) and time \(t\), \(f({\textbf{r}_s},t)\) is the source time function located at \({\textbf{r}_s}\), \(\delta \) denotes the Dirac delta function modeling the point source.

FWI relies on an objective function to evaluate the discrepancy between observed and simulated data. Here, we use the commonly adopted and widely regarded as the simplest Euclidean loss:
\begin{equation}
\label{eq2}
\Phi ({\bf{m}}) = {\left\| {{{\bf{d}}_{syn}}({\bf{m}}) - {{\bf{d}}_{obs}}} \right\|_2},
\end{equation}
where \({{{\bf{d}}_{syn}}}\) and \({{{\bf{d}}_{obs}}}\) denote the synthetic and observed data, respectively, \(\textbf{m}\) represents the subsurface velocity model, and \({\left\|  \cdot  \right\|_2}\) stands for the L2 norm. To minimize the data misfit between observed and simulated data, FWI typically updates the velocity model iteratively using the following gradient descent (GD) optimizer:
\begin{equation}
\label{eq3}
{{\bf{m}}_{i + 1}} = {{\bf{m}}_i} - \lambda \frac{{\partial \Phi }}{{\partial {{\bf{m}}_i}}},
\end{equation}
where \(\partial \Phi /\partial {{\bf{m}}_i}\) is the gradient with respect to the velocity model at iteration \textit{i}, while \(\lambda \) is the step length. The gradient is typically computed using the adjoint-state method \cite{plessix2006review}, which requires an explicit derivation of the gradient update formula. In recent years, with the rapid development of machine learning, automatic differentiation (AD) has gained popularity for gradient computation, including in FWI \cite{richardson_alan_2023}. AD automatically calculates gradients based on the wavefields during forward modeling and backward propagation, yielding results equivalent to the adjoint-state method. Moreover, it facilitates the seamless integration of machine learning techniques into FWI. In this study, we adopt the Deepwave toolbox proposed by \cite{richardson_alan_2023}, which formulates the forward modeling process as a recurrent neural network (RNN) and employs AD to compute velocity gradients \cite{richardson2018seismic, sun2020theory}.

Based on the GD optimizer, momentum is introduced to accelerate convergence and reduce oscillatory behavior during optimization \cite{qian1999momentum}. By incorporating a moving average of past gradients, it allows the optimizer to gain inertia in directions of consistent descent, thereby facilitating faster convergence. Momentum-based GD can be expressed by the following equations:

\begin{equation}
\label{eq4}
{c_i} = \beta {c_{i - 1}} + (1 - \beta )\frac{{\partial \Phi }}{{\partial {{\bf{m}}_i}}},
\end{equation}

\begin{equation}
\label{eq5}
{{\bf{m}}_{i + 1}} = {{\bf{m}}_i} - \lambda {c_i},
\end{equation}
where \({c_i}\) is the momentum term at iteration \textit{i},  \(\beta \) is the momentum coefficient.

The Adam optimizer \cite{kingma2014adam} is widely used in machine learning and FWI for its efficiency, robustness, and ability to adaptively adjust learning rates. It integrates the benefits of Momentum, which accelerates convergence by leveraging the accumulated history of gradients, and Root Mean Square Propagation (RMSProp), which adaptively tunes the learning rate for each parameter according to the recent magnitude of its gradients \cite{tieleman2012rmsprop}. Specifically, the Adam algorithm estimates both the first-order moment (the mean) and the second-order moment (the uncentered variance) of the gradients using exponentially decaying averages. The update process is governed by the following equations: 

\begin{equation}
\label{eq6}
{a_i} = {\beta _1}{a_{i - 1}} + (1 - {\beta _1})\frac{{\partial \Phi }}{{\partial {{\bf{m}}_i}}},
\end{equation}

\begin{equation}
\label{eq7}
{b_i} = {\beta _2}{b_{i - 1}} + (1 - {\beta _2}){\left( {\frac{{\partial \Phi }}{{\partial {{\bf{m}}_i}}}} \right)^2},
\end{equation}

\begin{equation}
\label{eq8}
{\hat a_i} = \frac{{{a_i}}}{{1 - \beta _1^i}},
\end{equation}

\begin{equation}
\label{eq9}
{\hat b_i} = \frac{{{b_i}}}{{1 - \beta _2^i}},
\end{equation}

\begin{equation}
\label{eq10}
{{\bf{m}}_{i + 1}} = {{\bf{m}}_i} - \lambda \frac{{{{\hat a}_i}}}{{\sqrt {{{\hat b}_i}}  + \varepsilon }},
\end{equation}
where \({a_i}\) is the first moment estimate (exponential moving average of the gradients), \({b_i}\) is the second moment estimate (exponential moving average of the squared gradients), \({\beta _1}\) is the decay rate for the first moment, \({\beta _2}\) is the decay rate for the second moment, \({\hat a_i}\) and \({\hat b_i}\) are the bias-corrected first and second moment estimates, and \(\varepsilon \) is a small positive constant to prevent division by zero. Adam's ability to adapt the learning rate individually for each parameter and to incorporate momentum makes it particularly effective for fast convergence and escaping local minima.

\section{NUMERICAL EXPERIMENTS}
In the field of machine learning, gradient descent or the Adam optimizer is often used to train neural network models with random initialization. With a relatively large learning rate, a well-trained model can be obtained often after thousands of iterations. The use of a large learning rate makes the optimization behave like a quasi-global algorithm. Motivated by this, we apply the same optimization strategy to FWI and find that it can effectively resolve cycle skipping. In this section, we use three synthetic tests and one field data test to demonstrate that FWI with gradient-based optimizers and a large step size can resolve cycle skipping given sufficient iterations. All seismic wave simulations in this study are performed using a finite-difference scheme to solve the acoustic wave equation (Equation \ref{eq1}), with second-order accuracy in time and eighth-order accuracy in space. A free-surface boundary condition is applied at the top to simulate multiple reflections, while perfectly matched layers (PML) conditions are used on the other boundaries to suppress artificial reflections \cite{berenger1994perfectly}. We remove frequency components below 3 Hz from the observed data in all synthetic tests.

\subsection{A linearly increasing model}
\begin{figure*}
\centering
\includegraphics[width=1\textwidth]{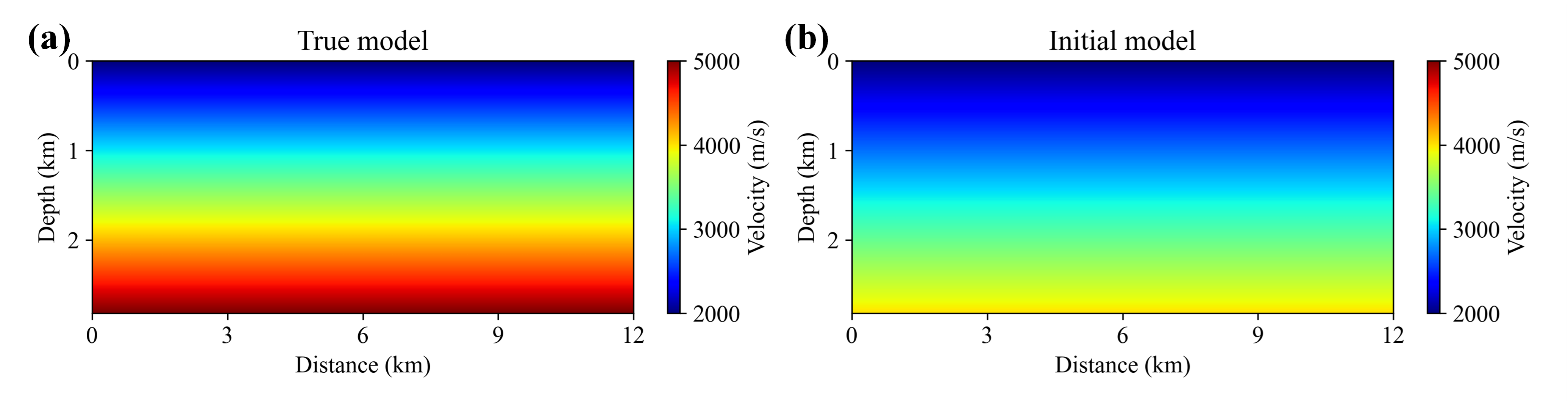}
\caption{(a) The true linearly increasing model and (b) the corresponding initial model used in FWI.}
\label{fig1}
\end{figure*}

We first use a simple linearly increasing velocity model, as shown in Fig. \ref{fig1}(a), to demonstrate that gradient-based optimizers can resolve cycle skipping and produce accurate inversion results given a sufficient iterations. The velocity model is defined on a grid of 400 × 94, with a uniform spacing of 30 m in both horizontal and vertical directions. A Ricker wavelet with a dominant frequency of 5 Hz is used as the source. We generate 30 shots, evenly spaced at 390 m intervals. Each shot is recorded by 400 receivers spaced 30 meters apart, with both sources and receivers located at the surface. The time sampling interval is 3 ms, and the total recording duration is 6 seconds. The simulated and observed shot gathers exhibit significant cycle skipping, as shown in Fig. \ref{fig4}(a). 

\begin{figure*}
\centering
\includegraphics[width=1\textwidth]{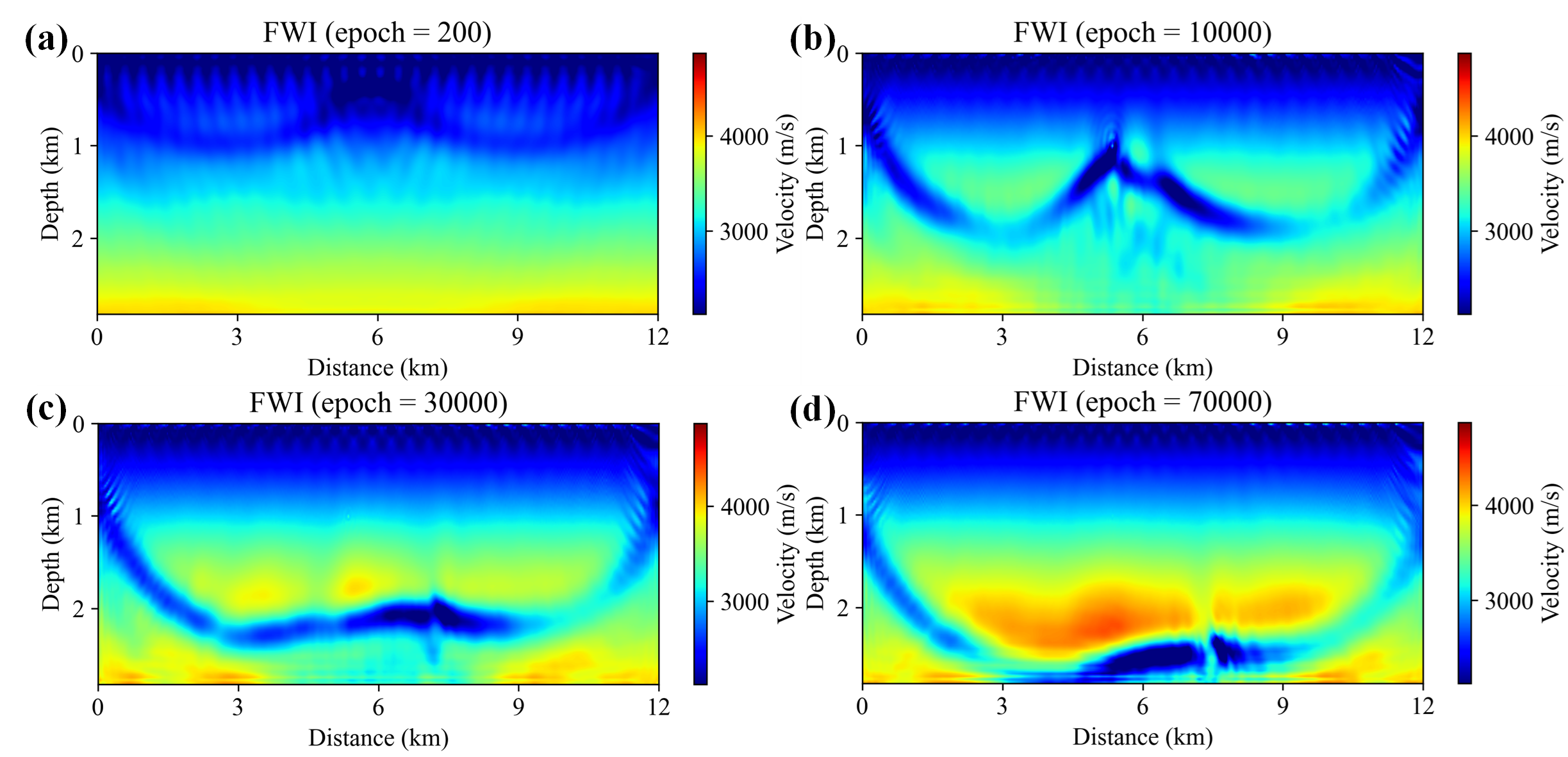}
\caption{FWI inversion results using the GD optimizer: (a)–(d) show the inverted results after 200, 10000, 30000, and 70000 iterations, respectively.}
\label{fig2}
\end{figure*}

\begin{figure*}
\centering
\includegraphics[width=1\textwidth]{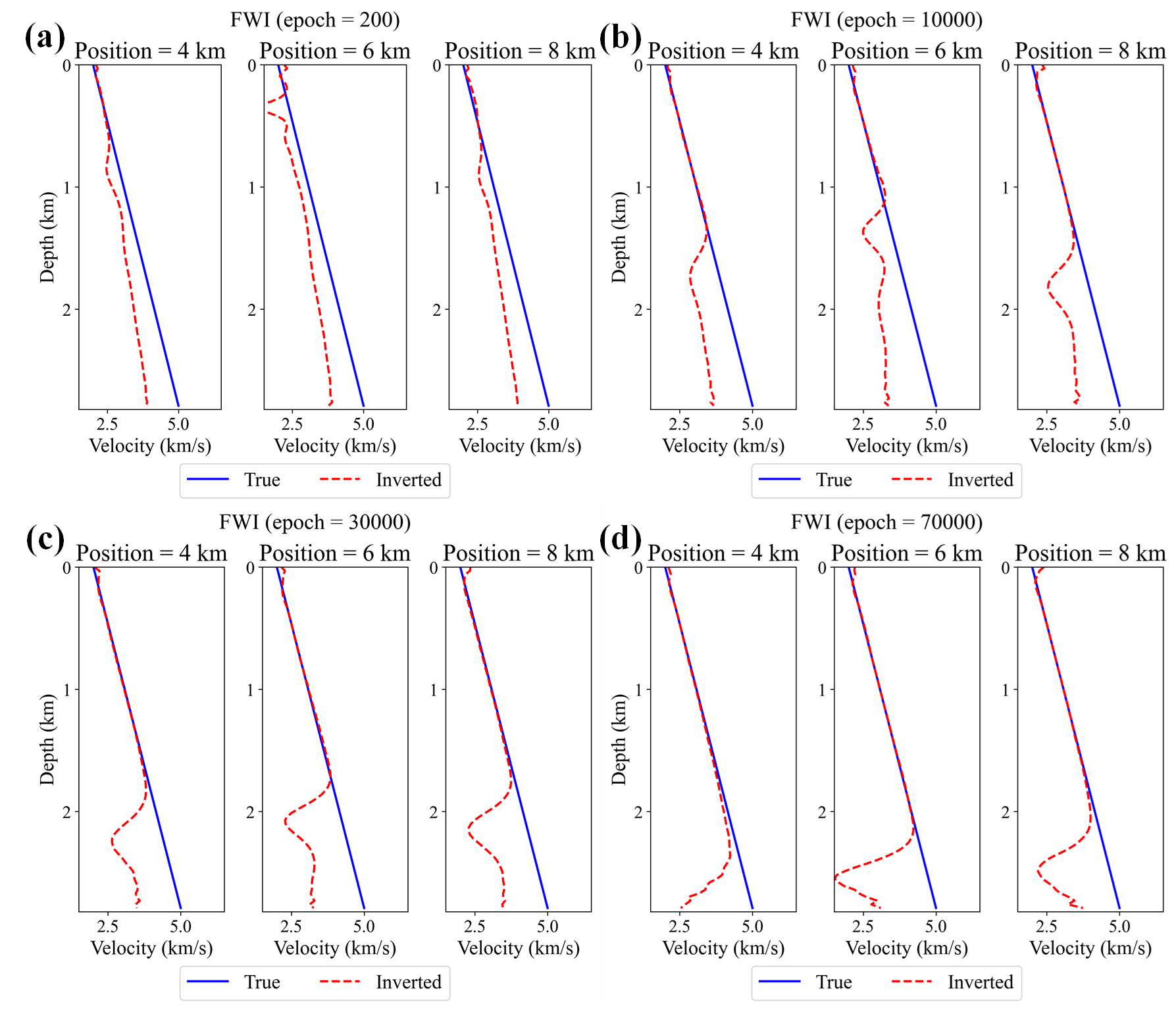}
\caption{One-dimensional velocity profiles extracted from horizontal locations 4 km, 6 km, and 8 km from Fig. \ref{fig2}. Panels (a)–(d) correspond to the profiles obtained from Figs \ref{fig2}(a)–\ref{fig2}(d), respectively. The green dashed line and the red dashed line represent the true velocity and the inverted velocity, respectively.}
\label{fig3}
\end{figure*}

\begin{figure*}
\centering
\includegraphics[width=0.7\textwidth]{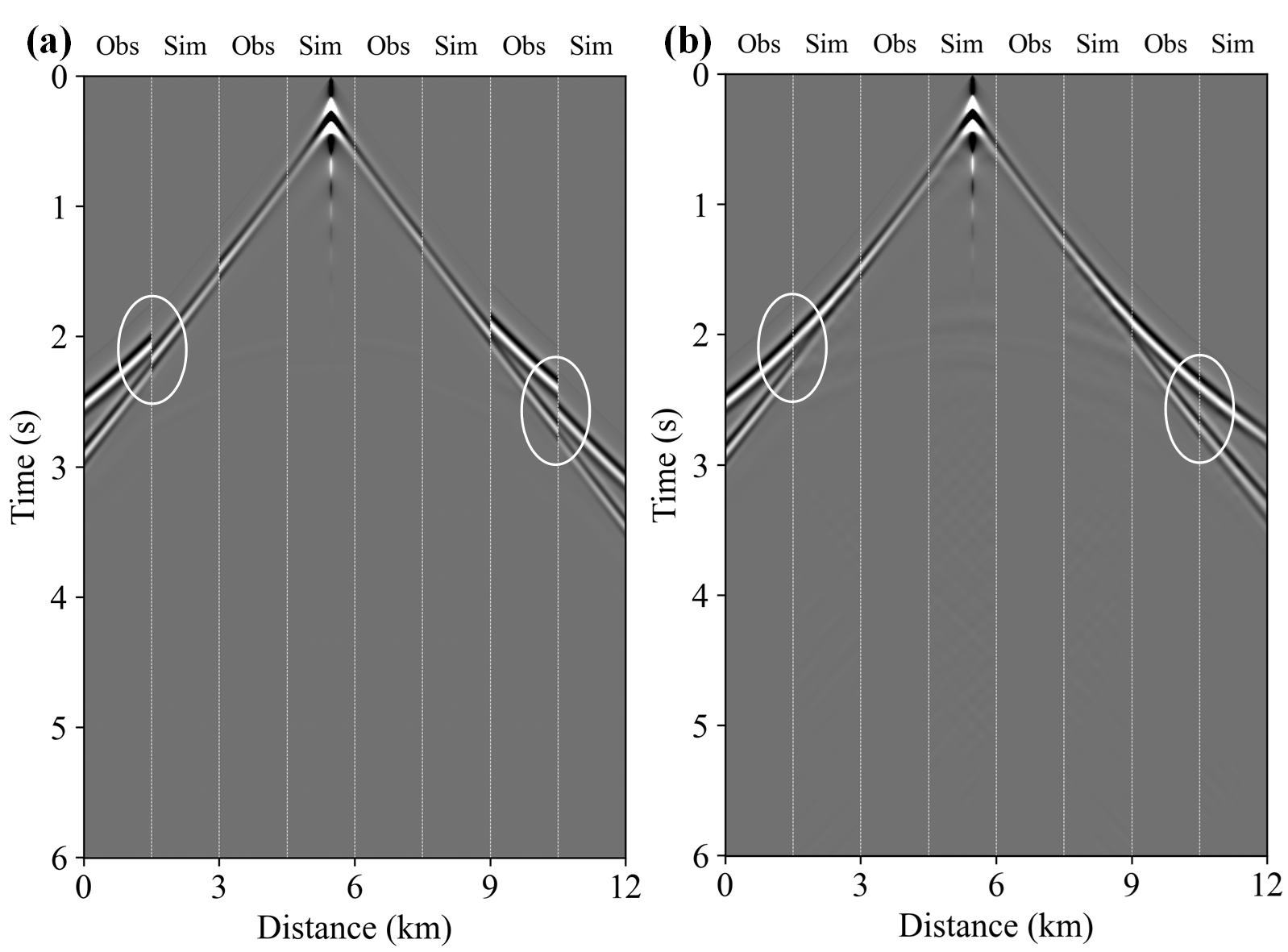}
\caption{Shot gather comparison. (a) and (b) display the 15th shot gather at iterations 0 and 70000, respectively, with 'Obs' and 'Sim' indicating the observed and simulated data.}
\label{fig4}
\end{figure*}

\begin{figure*}
\centering
\includegraphics[width=1\textwidth]{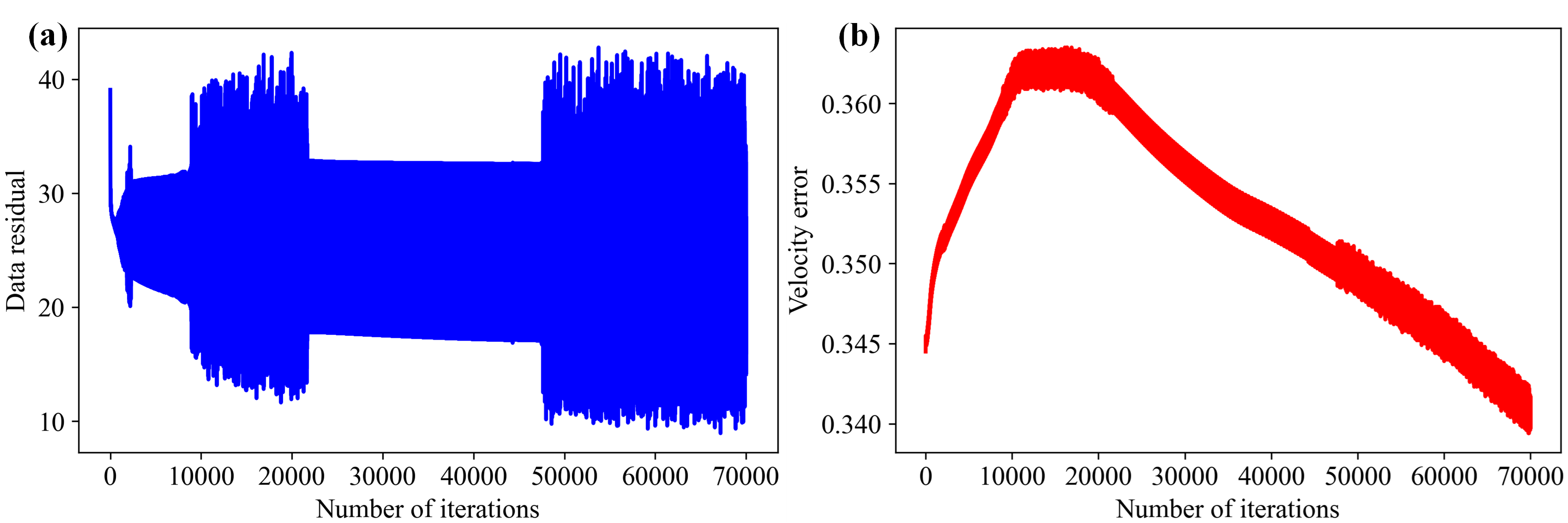}
\caption{Evolution of data residuals and velocity error curves as a function of iteration number for the linearly increasing velocity model test.}
\label{fig51}
\end{figure*}

Fig. \ref{fig2} shows the inversion results of FWI based on the GD optimizer after 200, 10000, 30000, and 70000 iterations. Fig. \ref{fig3} shows the 1D velocity profiles extracted from Fig. \ref{fig2} at horizontal positions of 4 km, 6 km, and 8 km, providing a more intuitive illustration of the inversion accuracy. The step length is set to 20000. As shown in Figs \ref{fig2}(a) and \ref{fig3}(a), there are clear differences between the inverted and true velocities after 200 iterations. As the number of iterations increases, FWI initially recovers accurate velocities in the shallow regions, followed by progressively improved inversion accuracy at greater depths. In other words, the inversion process gradually progresses from shallow to deep layers. This is because seismic information for surface recordings propagate from shallow to deep layers. Consequently, FWI must first achieve an accurate match between the observed and simulated data for reflections generated by shallow structures, followed by matching the seismic waves originating from deeper layers. Fig. \ref{fig4}(b) displays a comparison between the observed and the simulated shot gather generated from the inverted velocity model after 70000 iterations. The observed and simulated data show an excellent match, indicating that the FWI inversion result closely approximates the true velocity model. This test demonstrates that, given a sufficient number of iterations, the GD optimizer can solve cycle skipping and achieve accurate inversion results. As shown in Fig. \ref{fig4}(b), weak reflections can be observed in the simulated data. This phenomenon is primarily attributed to numerical discretization: the continuous velocity gradient is represented on a discrete grid as a sequence of small stepwise changes in velocity, which act as numerous thin impedance contrasts. The cumulative effect of these small contrasts generates faint scattered energy that appears as subtle reflections in the synthetic gathers. Fig. \ref{fig51} shows the convergence curves of data residuals and velocity errors. We observe that the data residuals exhibit significant fluctuations as the number of iterations increases. This is because, throughout the entire FWI inversion process, we use a large and fixed step size rather than determining it via line search, which adjusts the step size to ensure a continuous decrease in data misfit. As a result, the data misfit in our FWI exhibits noticeable oscillations. However, it is precisely this large learning rate that enables us to escape local minima and obtain an accurate velocity model. As shown in Fig. \ref{fig51}, although the data misfit becomes quite large after about 10000 iterations, the velocity error continues to decrease. Such oscillatory variations in the data misfit resemble the loss curves observed during neural network training, as illustrated in Fig. 7(a) of \cite{alkhalifah2021wavefield}. A detailed explanation of why gradient descent optimizers, a large learning rate, and a sufficient number of iterations can resolve cycle skipping is provided in the Discussion section.

\begin{figure*}
\centering
\includegraphics[width=1\textwidth]{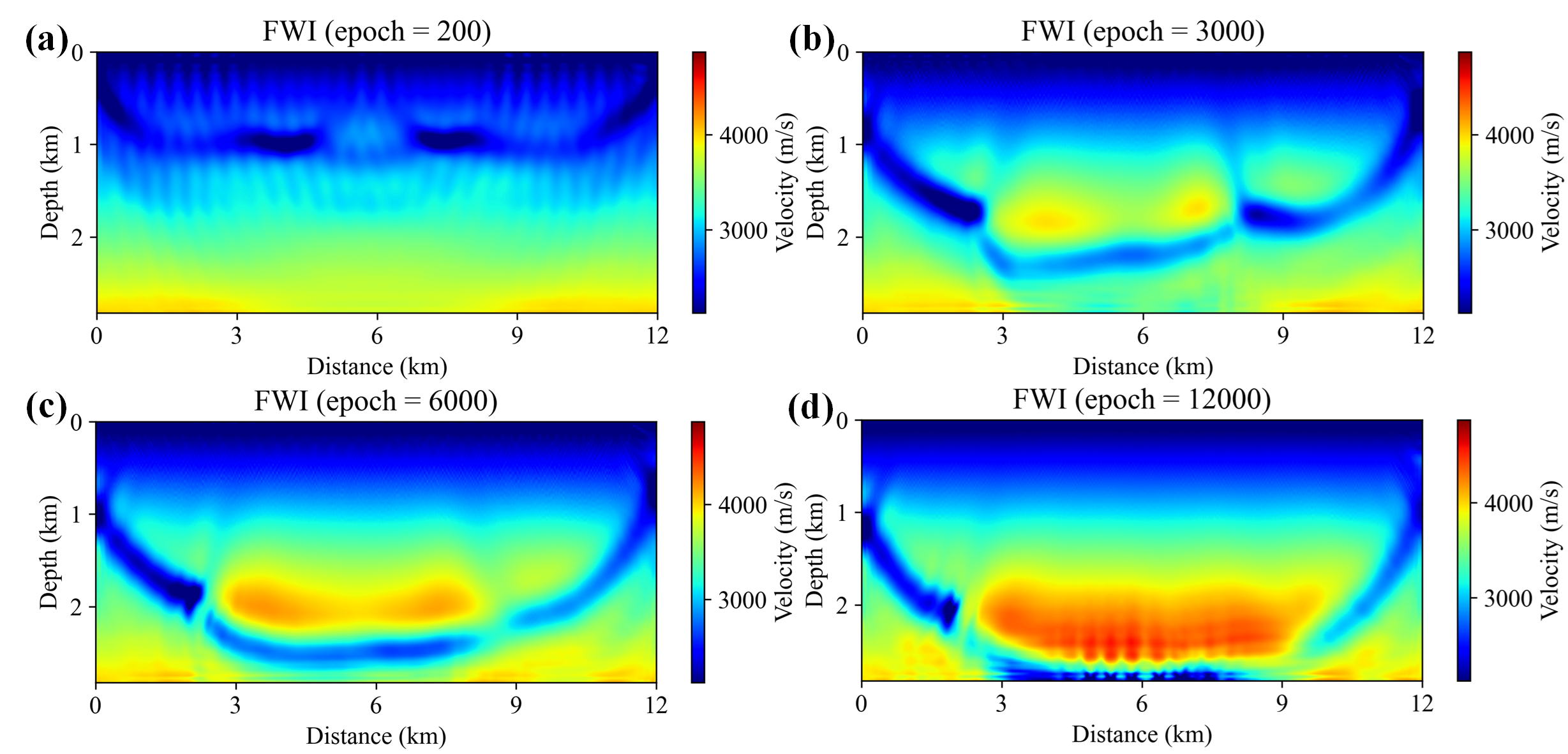}
\caption{FWI inversion results using the Momentum-based GD optimizer: (a)–(d) show the inverted results after 200, 3000, 6000, and 12000 iterations, respectively.}
\label{fig5}
\end{figure*}

\begin{figure*}
\centering
\includegraphics[width=1\textwidth]{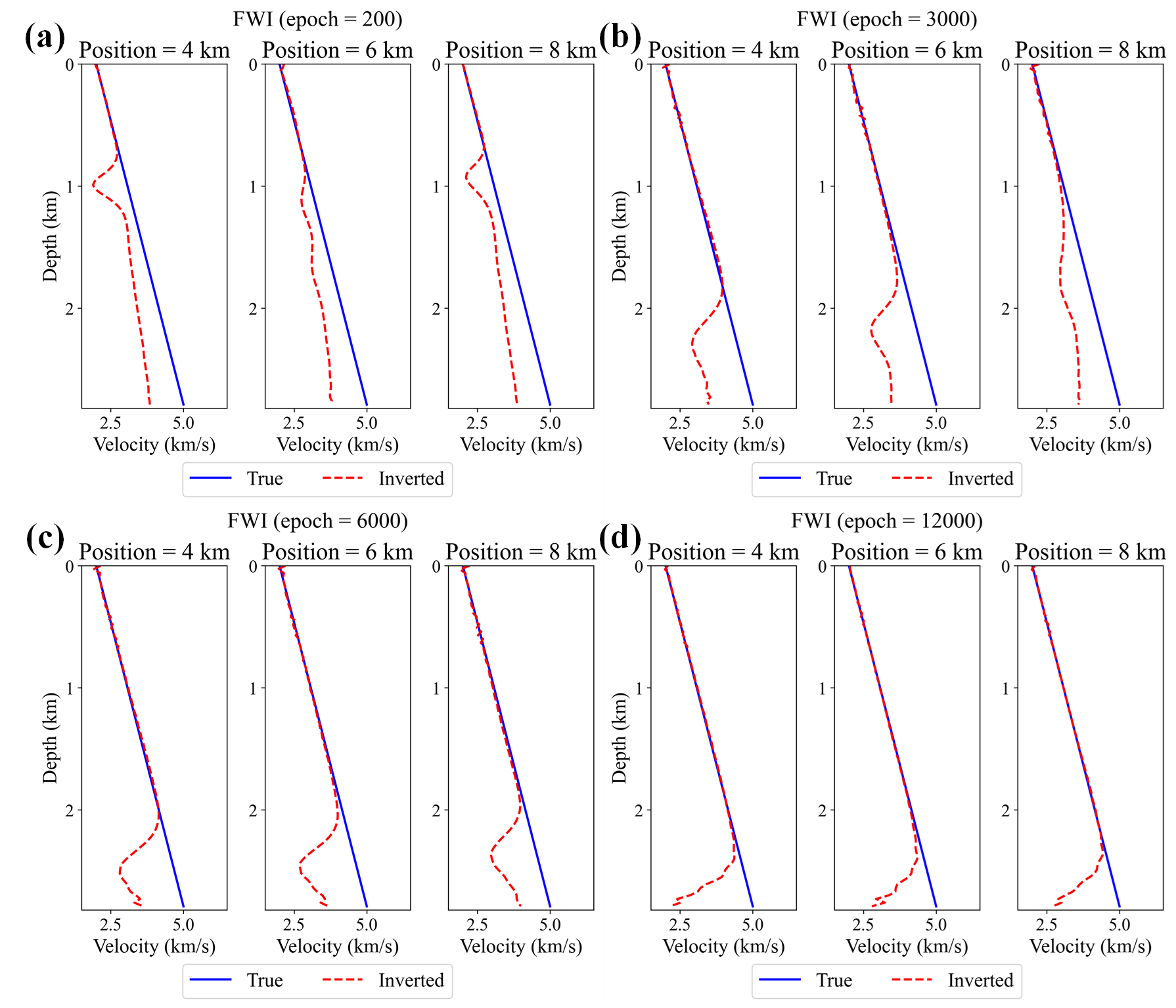}
\caption{One-dimensional velocity profiles extracted from horizontal locations 4 km, 6 km, and 8 km from Fig. \ref{fig5}. Panels (a)–(d) correspond to the profiles obtained from Figs \ref{fig5}(a)–\ref{fig5}(d), respectively.}
\label{fig6}
\end{figure*}

Furthermore, we perform FWI using a momentum-based GD optimizer, and the inversion results are shown in Fig. \ref{fig5}. The step length used in this test is 5000. Fig. \ref{fig6} shows 1D velocity profiles extracted from Fig. \ref{fig5}. Similarly, we observe that the velocity is updated progressively from shallow to deep regions. Although the inversion result after 200 iterations still shows a noticeable discrepancy from the true model, continued iterations gradually lead to a more accurate velocity reconstruction. Moreover, the momentum-based GD optimizer accelerates convergence toward the true velocity model compared to the standard GD optimizer. The result obtained after 12000 iterations using the momentum-based GD method (Figs \ref{fig5}(d) and \ref{fig6}(d)) is even better than that obtained after 70000 iterations using the standard GD optimizer (Figs \ref{fig2}(d) and \ref{fig3}(d)).

\begin{figure*}
\centering
\includegraphics[width=1\textwidth]{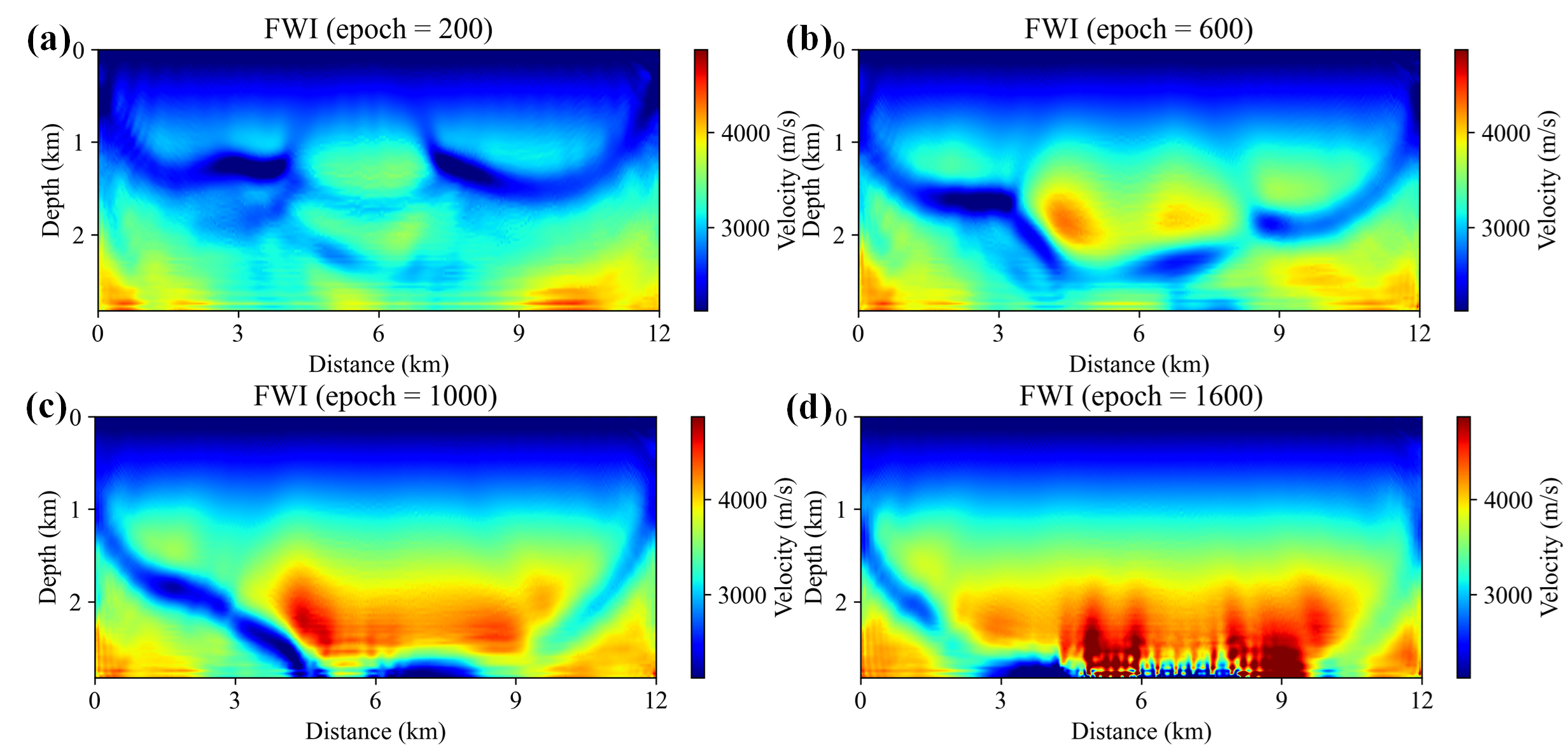}
\caption{FWI inversion results using the Adam optimizer: (a)–(d) show the inverted results after 200, 600, 1000, and 1600 iterations, respectively.}
\label{fig7}
\end{figure*}

\begin{figure*}
\centering
\includegraphics[width=1\textwidth]{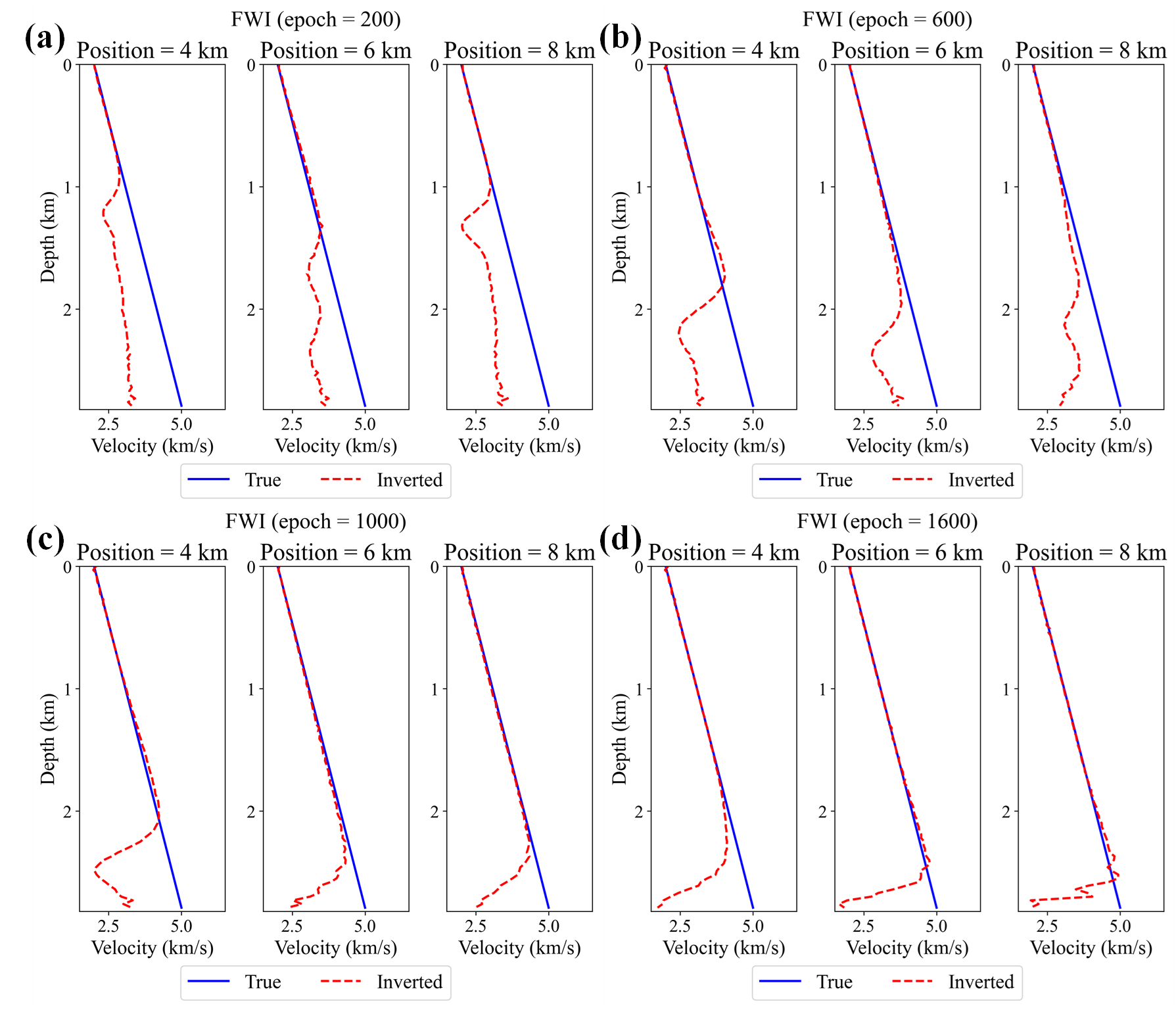}
\caption{One-dimensional velocity profiles extracted from horizontal locations 4 km, 6 km, and 8 km from Fig. \ref{fig7}. Panels (a)–(d) correspond to the profiles obtained from Figs \ref{fig7}(a)–\ref{fig7}(d), respectively.}
\label{fig8}
\end{figure*}

We further perform FWI using the Adam optimizer, and the inverted models are shown in Fig. \ref{fig7}. Similarly, Fig. \ref{fig8} shows the 1D velocity profiles extracted from Fig. \ref{fig7}. The step length is 50. Compared to the results obtained using the momentum-based GD optimizer (Fig. \ref{fig5}), FWI with the Adam optimizer obtains accurate inversion results with only 1600 iterations, significantly reducing the number of iterations required.

\begin{figure*}
\centering
\includegraphics[width=1\textwidth]{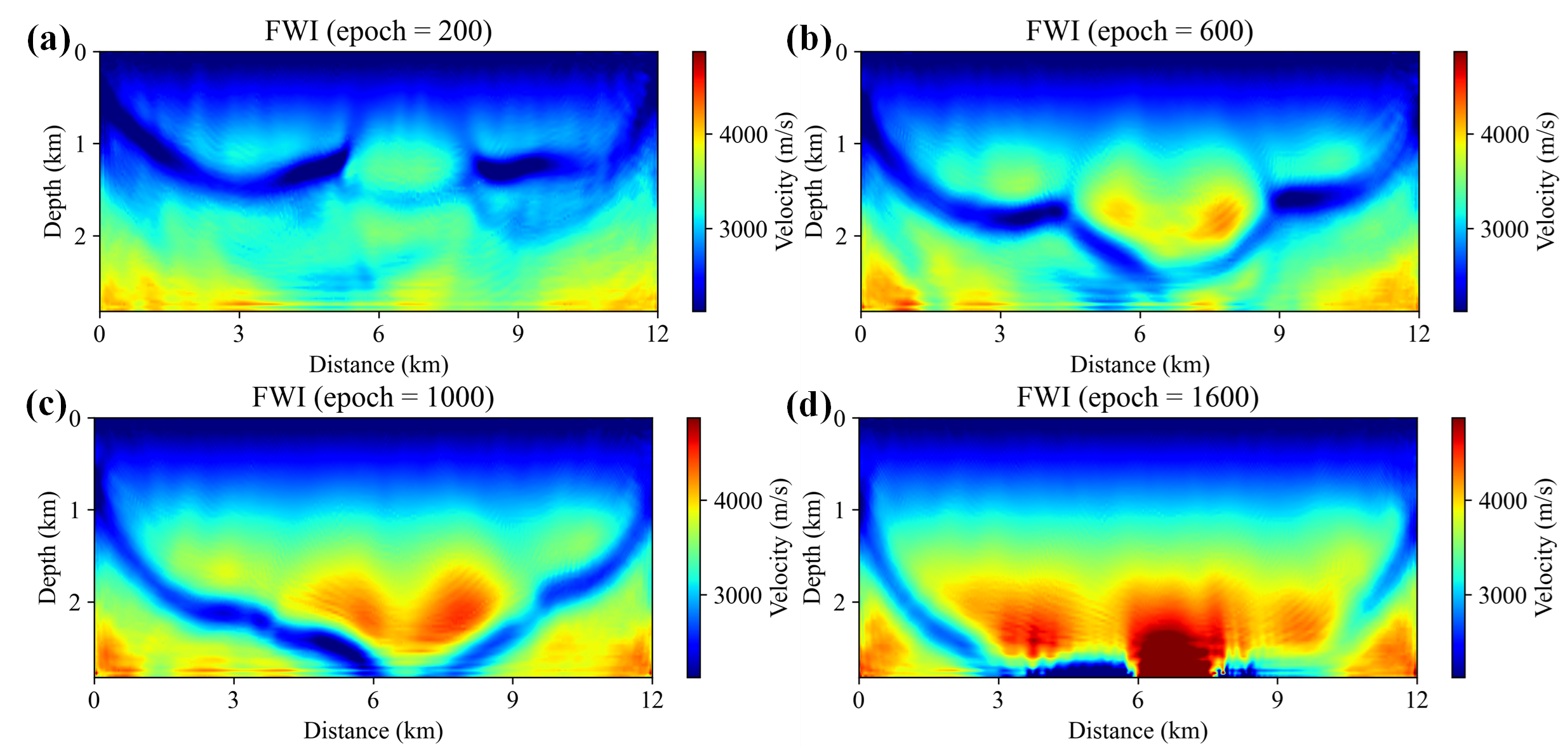}
\caption{FWI results with the mini-batch Adam optimizer: (a)–(d) display the inversion outputs after 200, 600, 1000, and 1600 iterations.}
\label{fig9}
\end{figure*}

\begin{figure*}
\centering
\includegraphics[width=1\textwidth]{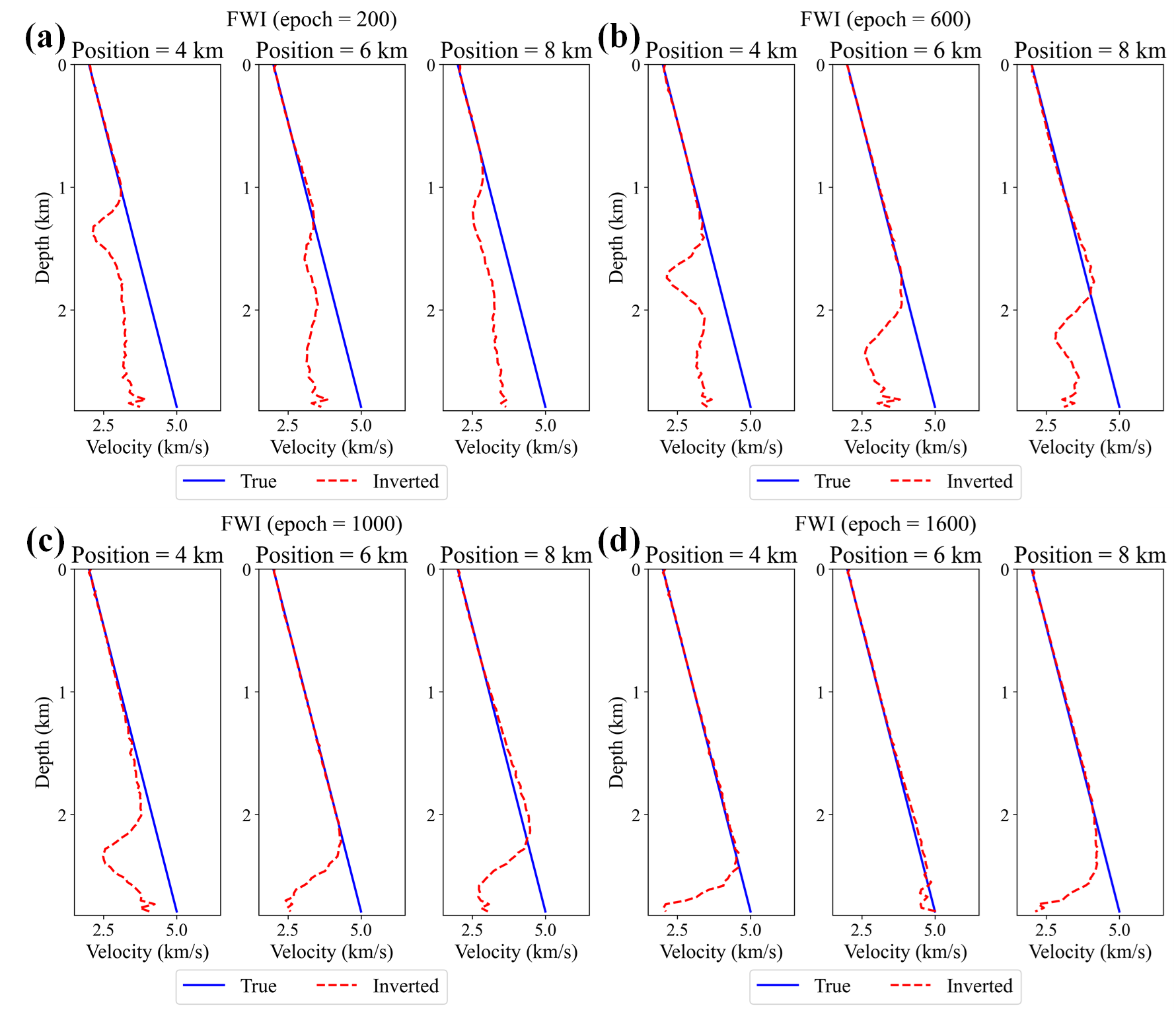}
\caption{One-dimensional velocity profiles at horizontal positions of 4 km, 6 km, and 8 km are extracted from Fig. \ref{fig9}. Panels (a)–(d) display the corresponding profiles from Figs \ref{fig9}(a)–\ref{fig9}(d), respectively.}
\label{fig10}
\end{figure*}

To further reduce the computational cost, we employ a mini-batch Adam optimizer, where only a subset of the shot gathers is used for gradient computation and velocity update in each iteration. Since the primary computational burden in FWI arises from wavefield simulations, this mini-batch strategy substantially reduces the computational cost of wavefield simulations per iteration, thereby improving overall efficiency. In this test, only half of the shot gathers are randomly selected for gradient computation in each iteration, resulting in a computational cost that is approximately half of that incurred by FWI using the Adam optimizer with all shots. Fig. \ref{fig9} shows the inversion results obtained using the mini-batch Adam optimizer, and Fig. \ref{fig10} shows single-trace comparisons extracted from Fig. \ref{fig9}. The step length is the same as that used in Fig. \ref{fig7}. Based on the above tests, we find that gradient-based optimizers, with a relatively large and intelligently picked step length (learning rate), can resolve cycle skipping; as the number of iterations increases, the shallow velocities are inverted accurately first, followed by the deeper velocities. We also observe that the mini-batch Adam optimizer significantly reduces computational cost compared to the GD optimizer. 

\subsection{Overthrust model}
\begin{figure*}
\centering
\includegraphics[width=1\textwidth]{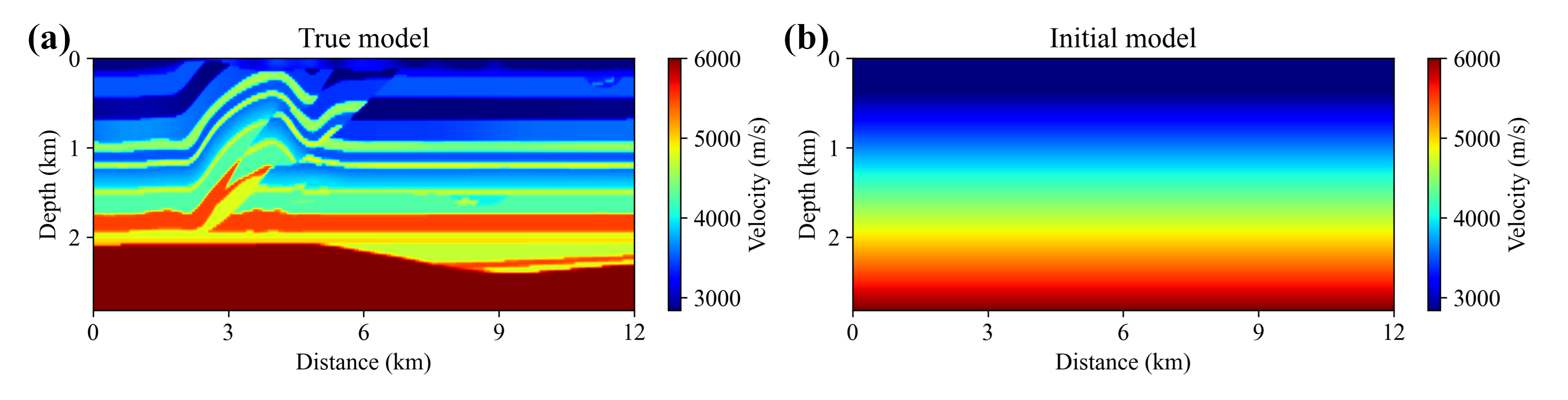}
\caption{(a) The true Overthrust model and (b) its starting model used for FWI.}
\label{fig11}
\end{figure*}

We use a complex Overthrust model \cite{aminzadeh1994seg} to validate that gradient-based optimizers, given a large number of iterations, can resolve cycle skipping. The true velocity model is shown in Fig. \ref{fig11}(a). The Overthrust model is discretized on a 400 × 94 grid with uniform spatial intervals of 30 meters in both horizontal and vertical directions. A Ricker wavelet with a peak frequency of 5 Hz is adopted as the source wavelet. Thirty shots are deployed uniformly along the surface at intervals of 390 meters, with the first source located at the leftmost boundary of the model domain. Each shot is recorded by an array of 400 receivers, equally spaced at 30 meters and positioned at the surface. The total recording time is set to 6 seconds per shot, with a temporal sampling interval of 3 ms. Fig. \ref{fig11}(b) shows the initial velocity model used for FWI, which is a linearly increasing model from top to bottom. 

\begin{figure*}
\centering
\includegraphics[width=1\textwidth]{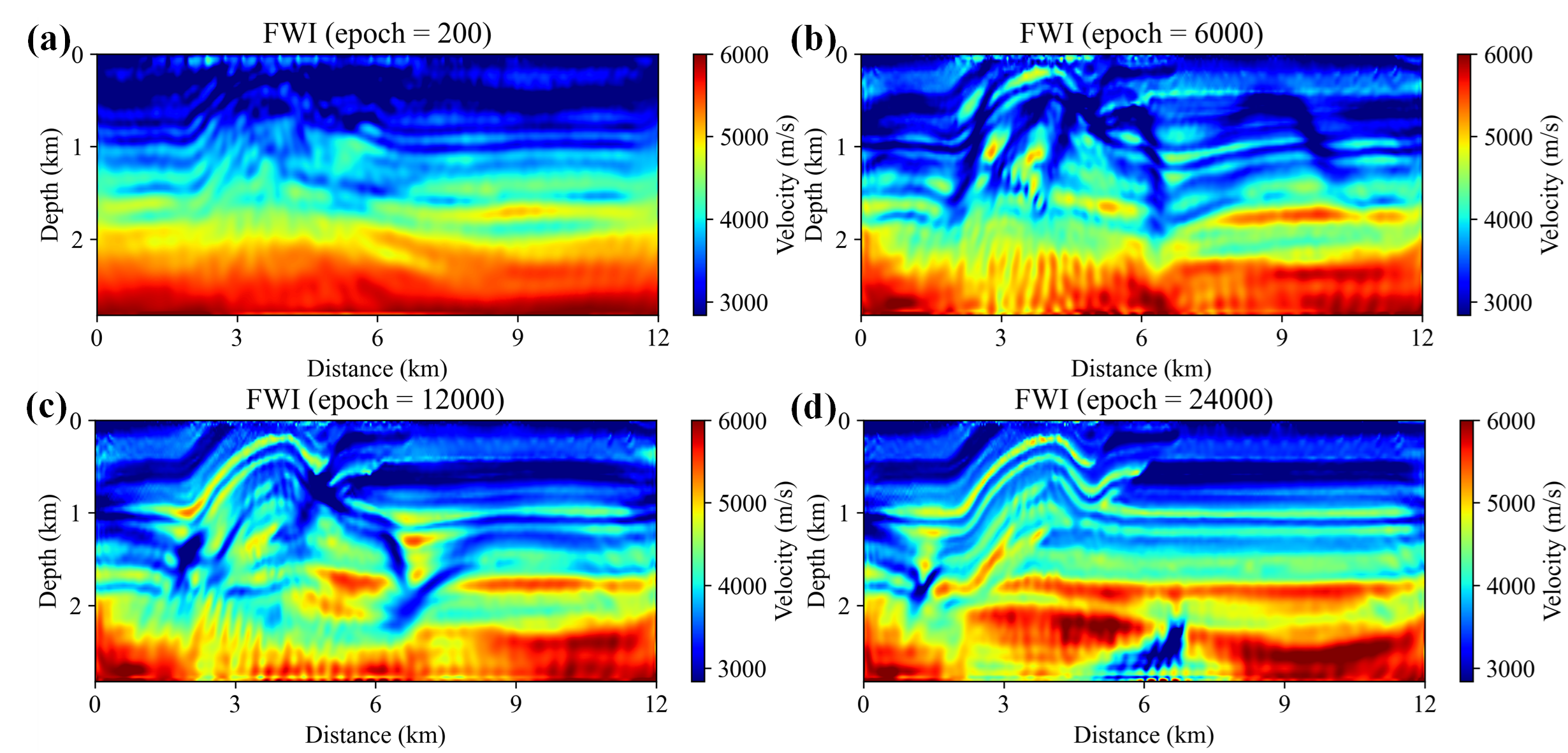}
\caption{Panels (a)–(d) show the FWI inversion results using the GD optimizer after 200, 6000, 12000, and 24000 iterations, respectively.}
\label{fig12}
\end{figure*}

\begin{figure*}
\centering
\includegraphics[width=1\textwidth]{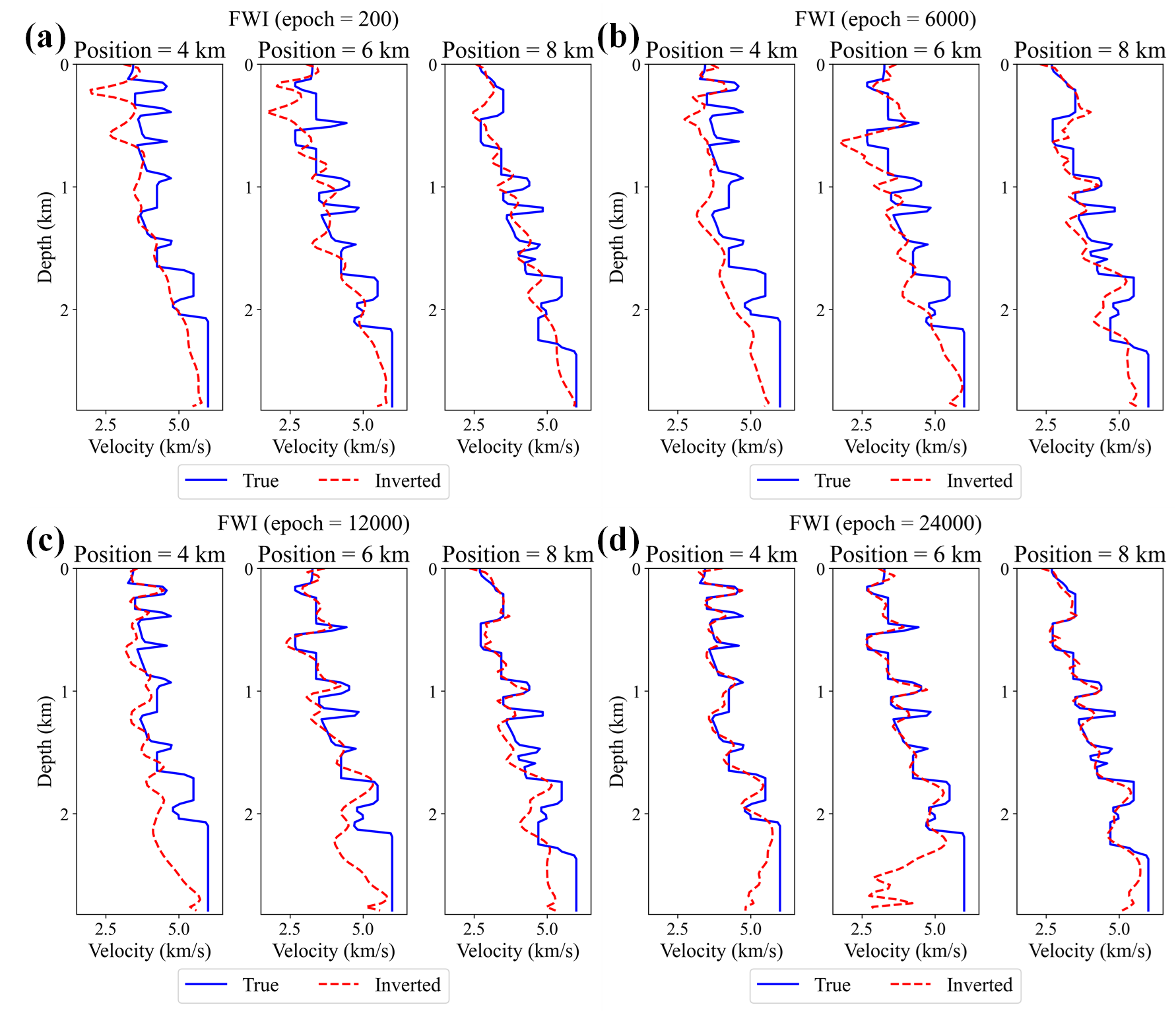}
\caption{Single-trace comparisons at horizontal distances of 4 km, 6 km, and 8 km are taken from Fig. \ref{fig12}, where panels (a)–(d) correspond to the velocity models in Figs \ref{fig12}(a)–\ref{fig12}(d), respectively.}
\label{fig13}
\end{figure*}

\begin{figure*}
\centering
\includegraphics[width=0.7\textwidth]{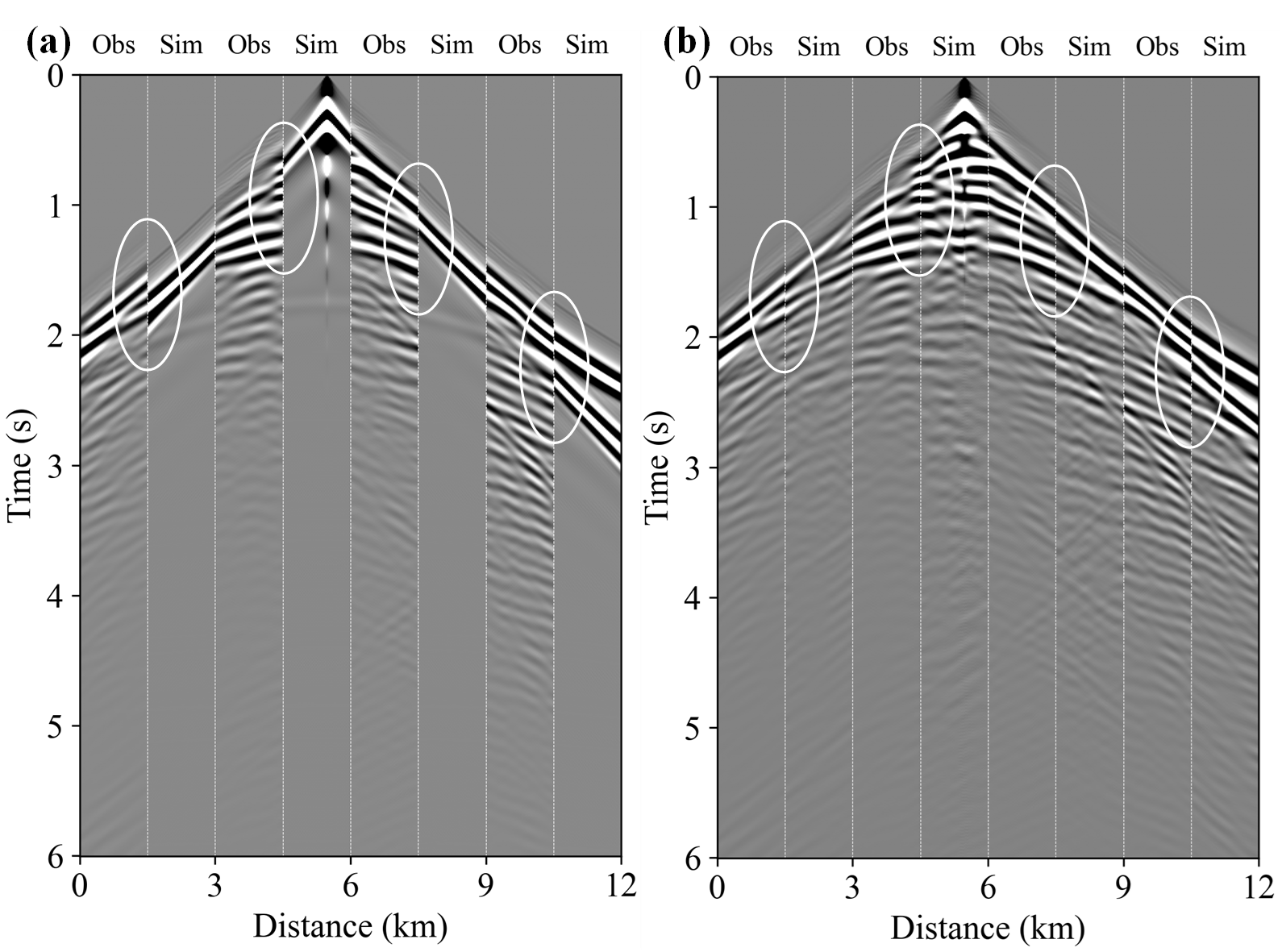}
\caption{Panels (a) and (b) show shot gather comparisons for the 15th shot, where the simulated data are computed using the inverted velocity models after 0 and 24000 iterations, respectively. The panels marked 'Obs' and 'Sim' indicate the observed and simulated data, respectively.}
\label{fig14}
\end{figure*}

We perform FWI using the GD optimizer, and the resulting inversion results are shown in Fig. \ref{fig12}. Fig. \ref{fig13} shows single-trace comparisons between the inverted velocities extracted from Fig. \ref{fig12} and the true velocities. The step length for this test is 100000. After 200 iterations, the inverted velocity exhibits significant discrepancies compared to the true velocity, as shown in Figs \ref{fig12}(a) and \ref{fig13}(a). With an increasing number of iterations, the inversion results progressively approach the true velocity. Similarly, accurate recovery is first achieved in the shallow regions, followed by deeper areas as the inversion proceeds. Figs \ref{fig14}(a) and \ref{fig14}(b) compare observed shot gathers with those simulated using the initial velocity model and the inverted velocity model after 24000 iterations, respectively. Fig. \ref{fig14}(a) shows clear evidence of cycle skipping between the observed and simulated data, as highlighted by the white ellipses. In contrast, Fig. \ref{fig14}(b) demonstrates an almost perfect match between the observed and simulated shot gathers, indicating that the velocity model inverted using the GD optimizer closely approximates the true velocity.

\begin{figure*}
\centering
\includegraphics[width=1\textwidth]{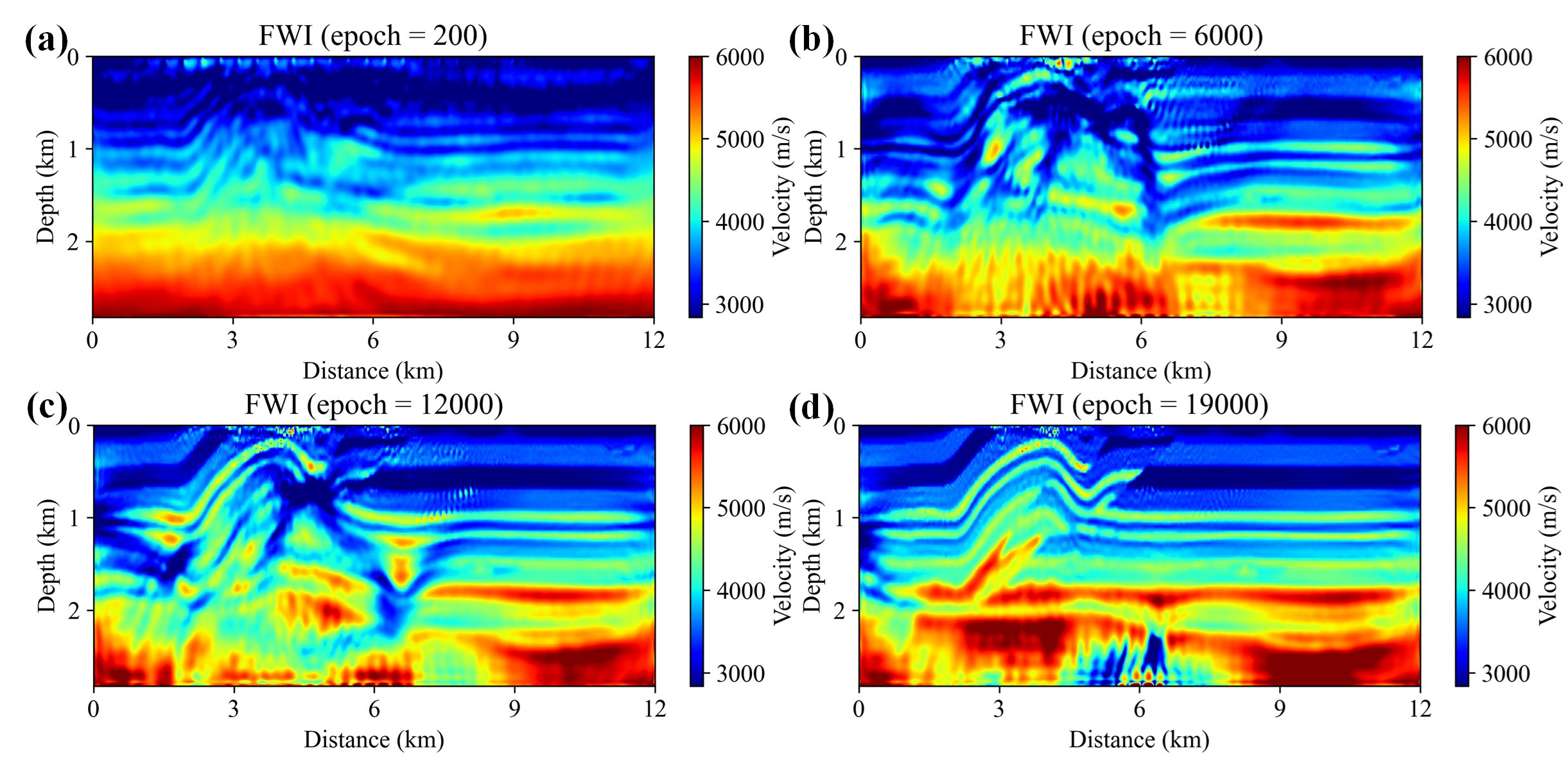}
\caption{Panels (a)–(d) show the FWI inversion results using the momentum-based GD optimizer after 200, 6000, 12000, and 19000 iterations, respectively.}
\label{fig15}
\end{figure*}

\begin{figure*}
\centering
\includegraphics[width=1\textwidth]{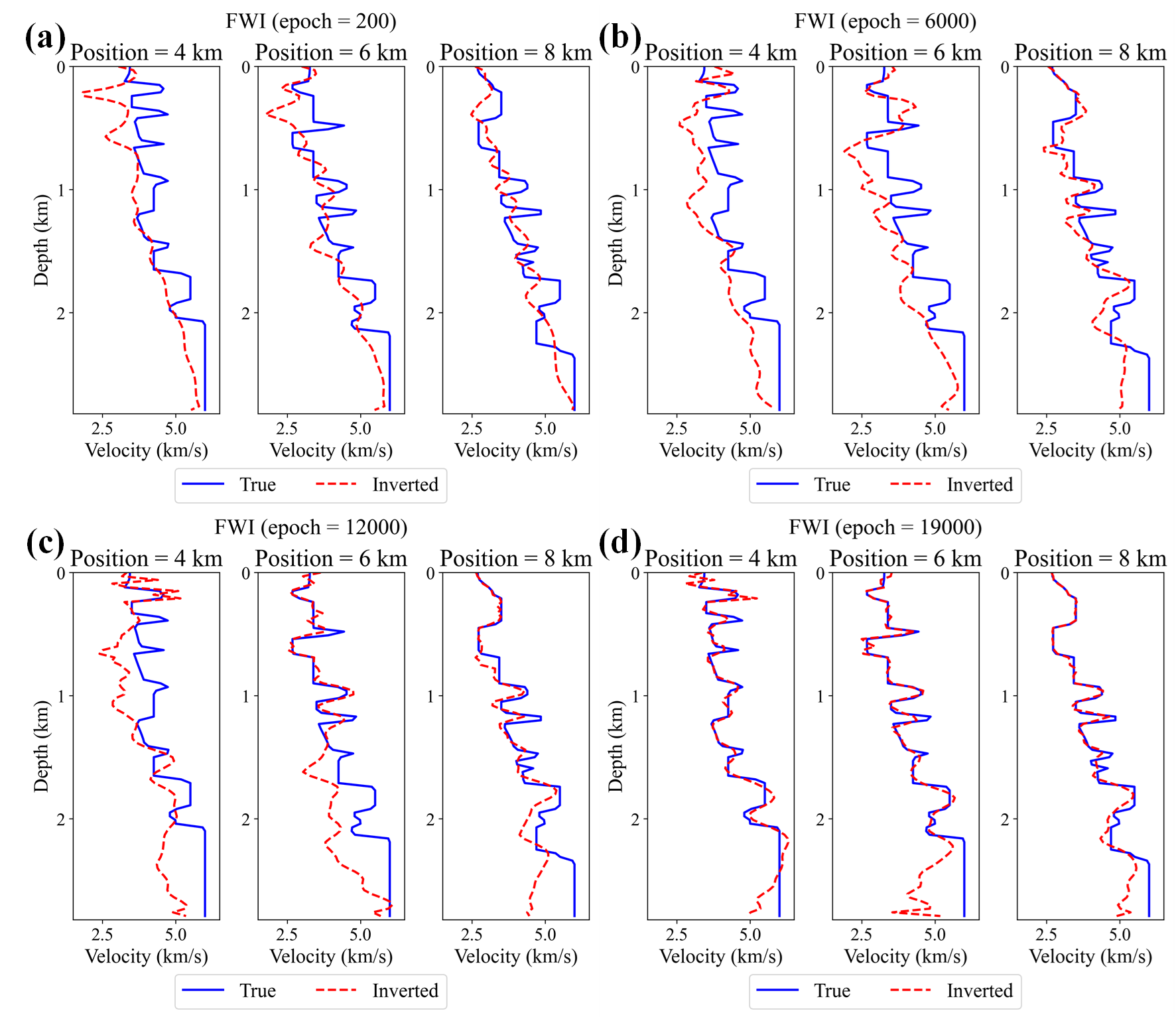}
\caption{Single-trace comparisons at horizontal distances of 4 km, 6 km, and 8 km are extracted from Fig. \ref{fig15}. Panels (a)–(d) correspond to the velocity profiles shown in Figs \ref{fig15}(a)–\ref{fig15}(d), respectively.}
\label{fig16}
\end{figure*}

The inversion result obtained using the momentum-based GD optimizer is shown in Fig. \ref{fig15}, with the step length set to 10000 for this test. Fig. \ref{fig16} shows the single-trace comparison extracted from Fig. \ref{fig15}. Compared to Fig. \ref{fig13}(d), the red and green lines in Fig. \ref{fig16} exhibit a closer match, indicating that the momentum-based GD optimizer produces more accurate inversion results with fewer iterations than the standard GD optimizer. This highlights the momentum-based GD optimizer's superior ability to overcome local minima and yield more accurate inversion results.

\begin{figure*}
\centering
\includegraphics[width=1\textwidth]{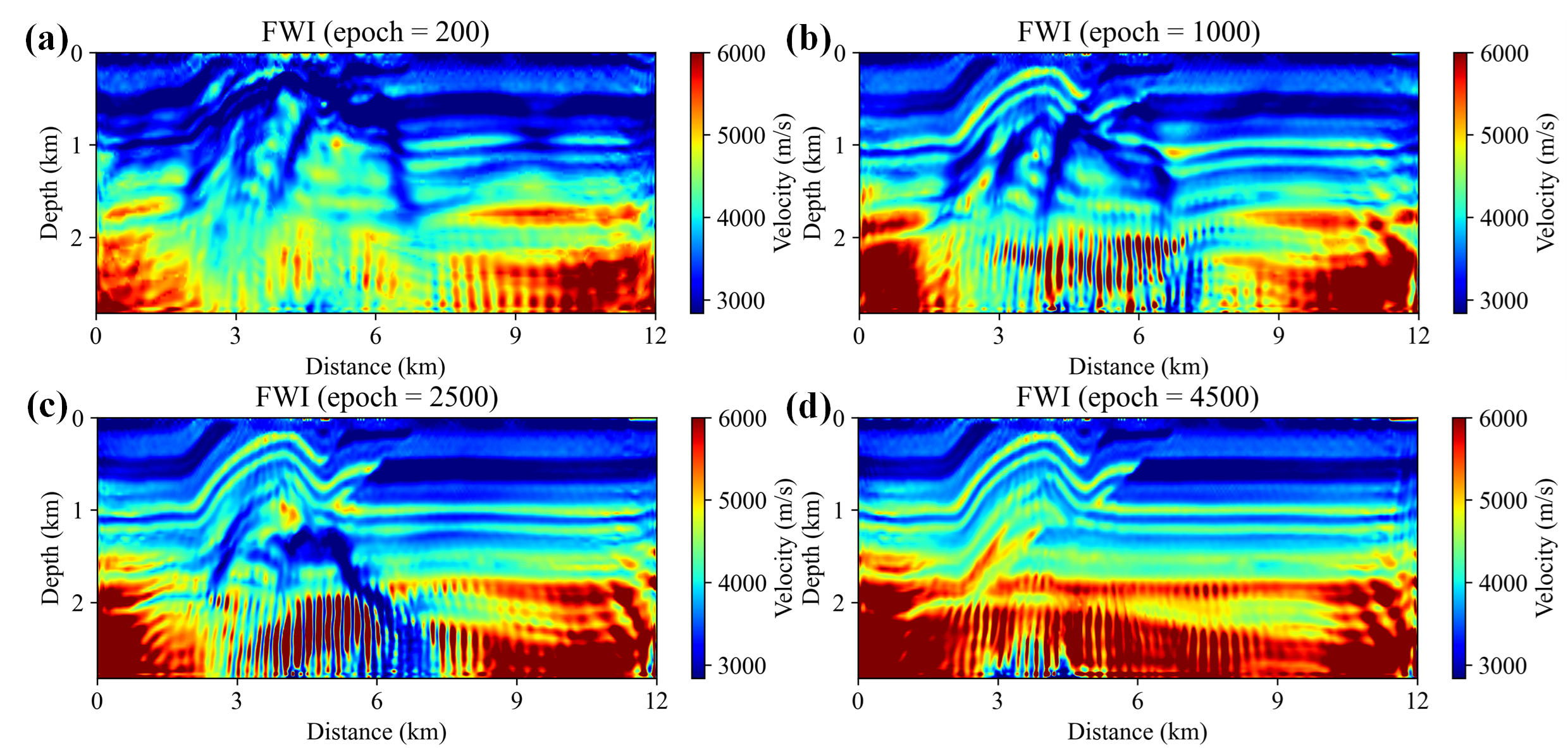}
\caption{Panels (a)–(d) show the FWI inversion results using the Adam optimizer after 200, 1000, 2500, and 4500 iterations, respectively.}
\label{fig17}
\end{figure*}

\begin{figure*}
\centering
\includegraphics[width=1\textwidth]{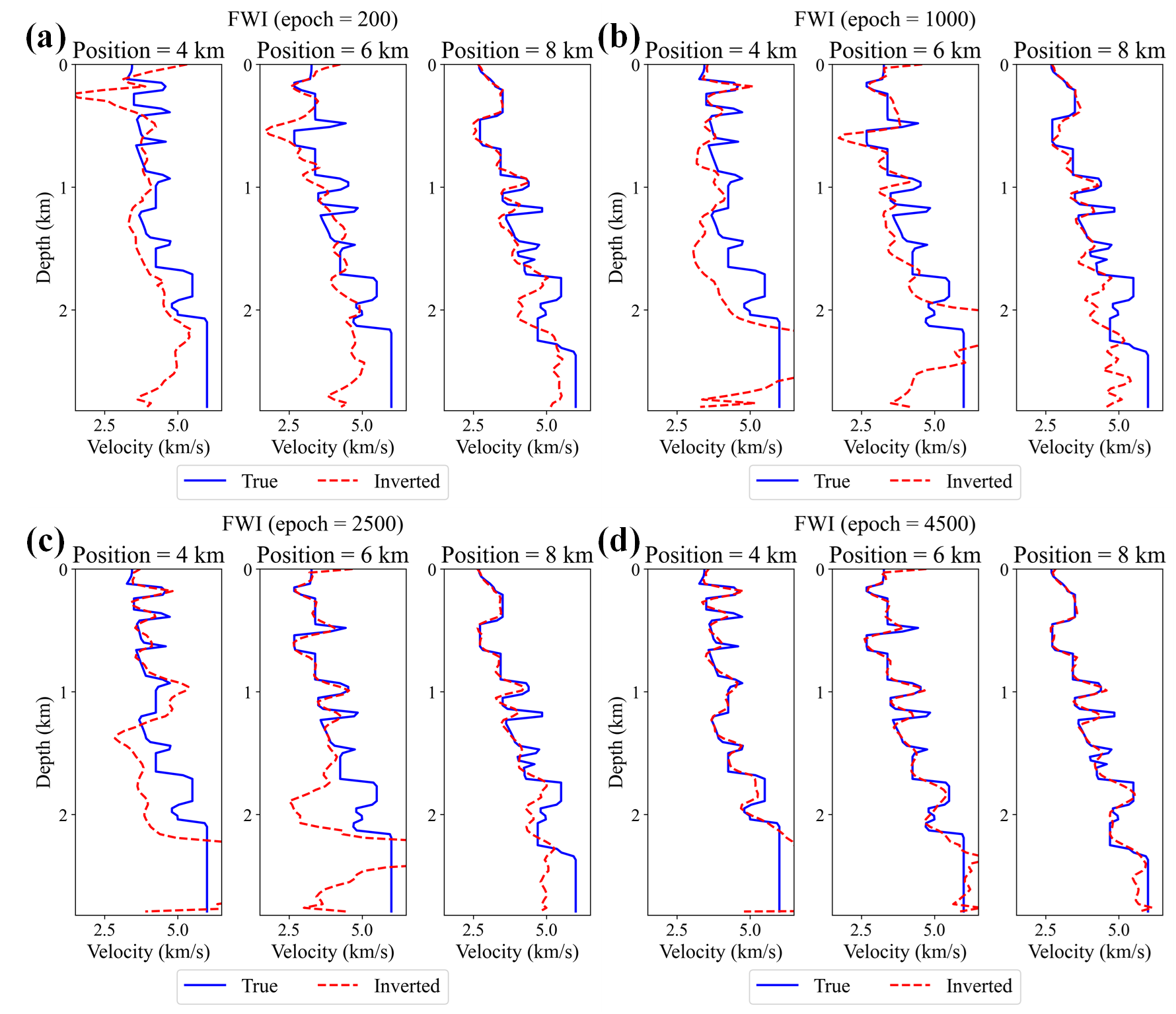}
\caption{Single-trace comparisons at horizontal distances of 4 km, 6 km, and 8 km are extracted from Fig. \ref{fig17}. Panels (a)–(d) correspond to the velocity profiles shown in Figs \ref{fig17}(a)–\ref{fig17}(d), respectively.}
\label{fig18}
\end{figure*}

We further employ the Adam optimizer for FWI inversion, and the inverted models are shown in Fig.\ref{fig17}. Single-trace comparisons extracted from Fig.\ref{fig17} is shown in Fig.\ref{fig18}. The step length used in this test is set to 150. Compared to the results obtained using the momentum-based GD optimizer (Figs \ref{fig15} and \ref{fig16}), the Adam optimizer obtains accurate inversion results with only 4500 iterations, whereas the momentum-based GD optimizer requires 19000 iterations to reach a similar level of accuracy. Therefore, the Adam optimizer significantly reduces the number of iterations required to achieve accurate velocity inversion compared to the momentum-based GD optimizer.

\begin{figure*}
\centering
\includegraphics[width=1\textwidth]{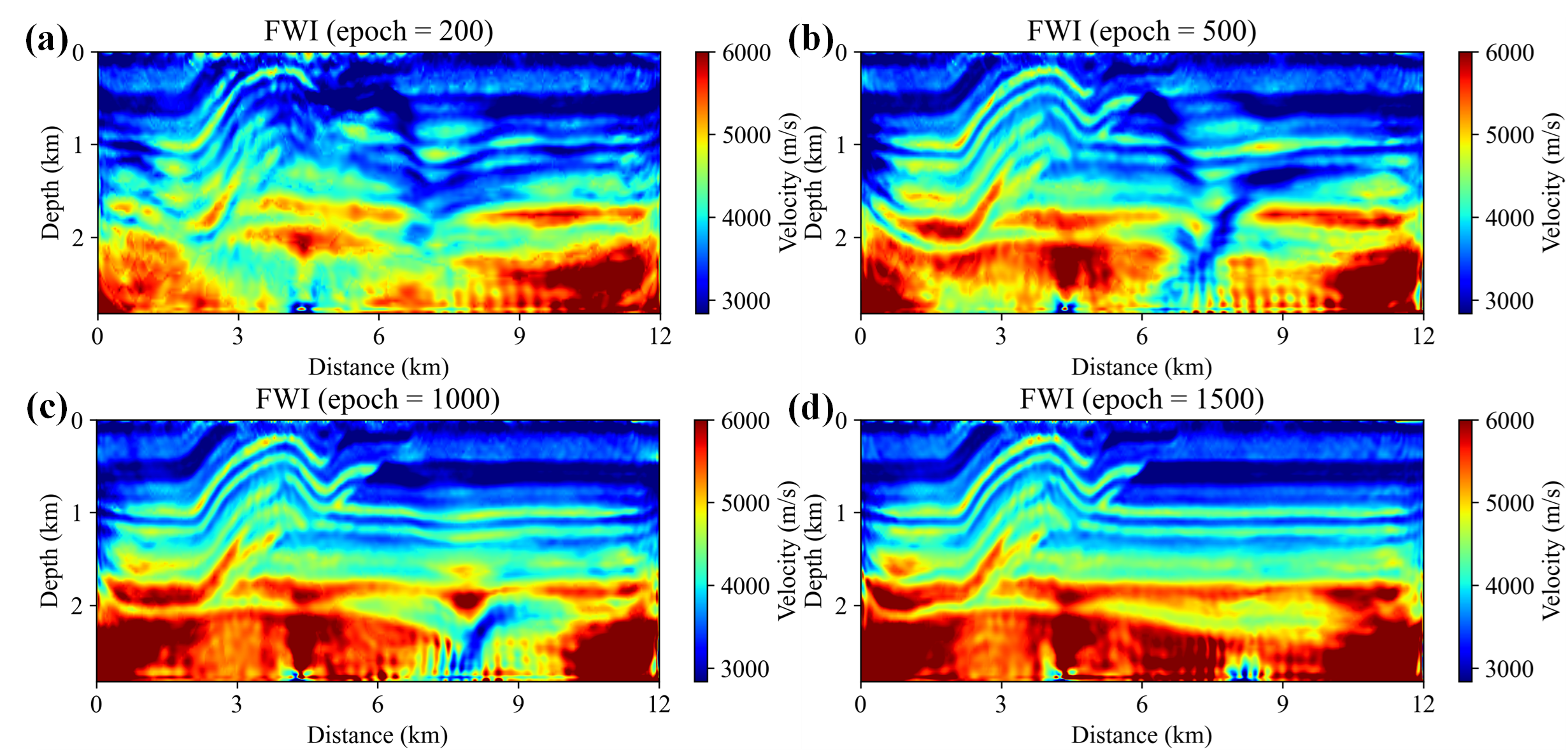}
\caption{FWI inversion results obtained using the mini-batch Adam optimizer. (a)–(d) correspond to 200, 500, 1000, and 1500 iterations, respectively.}
\label{fig19}
\end{figure*}

\begin{figure*}
\centering
\includegraphics[width=1\textwidth]{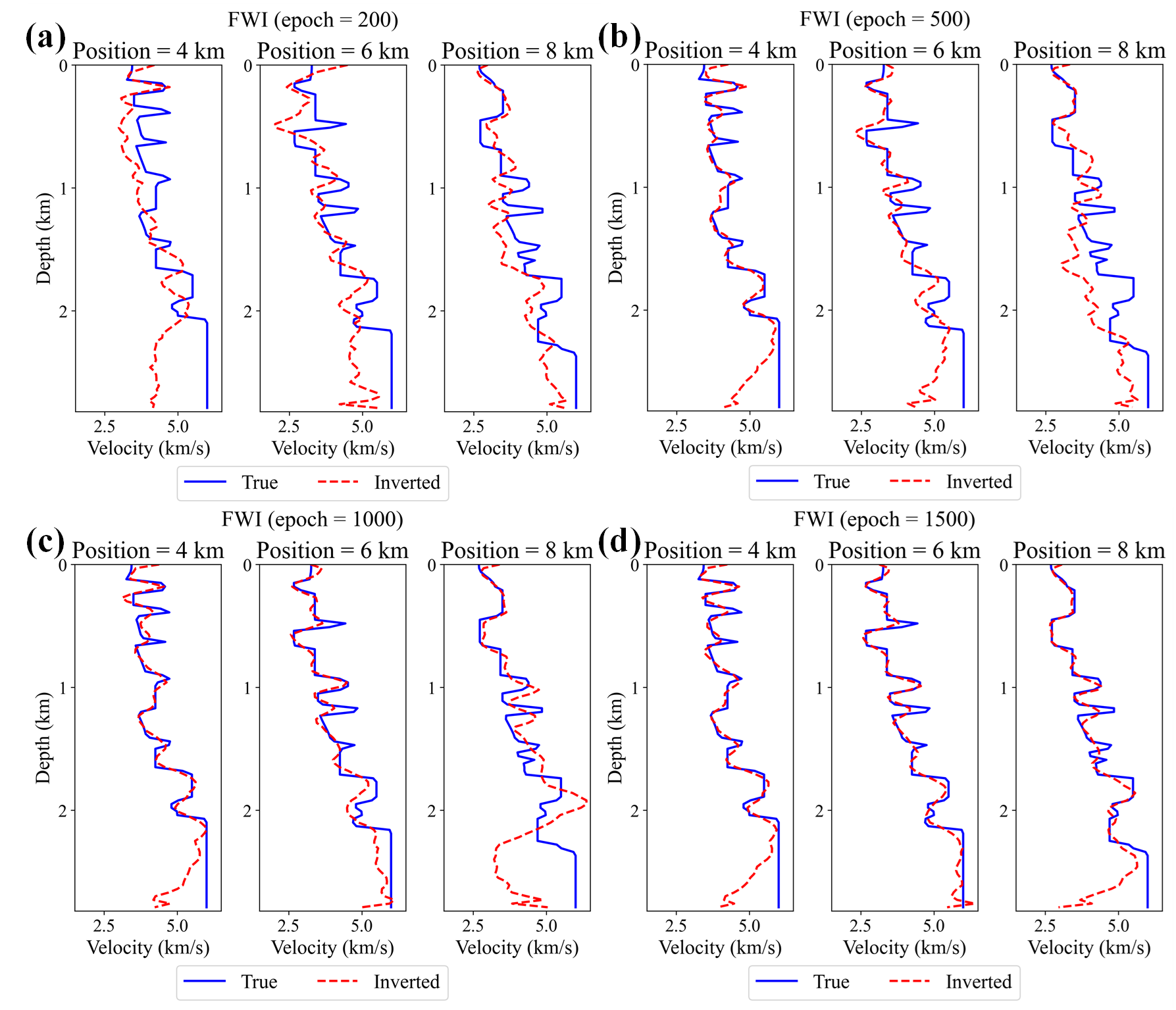}
\caption{Single-trace comparisons at horizontal distances of 4 km, 6 km, and 8 km are extracted from Fig. \ref{fig19}. Panels (a)–(d) correspond to the velocity profiles shown in Figs \ref{fig19}(a)–\ref{fig19}(d), respectively.}
\label{fig20}
\end{figure*}

We consider incorporating the mini-batch strategy to further reduce computational cost. Here, only 10 shots are randomly selected in each iteration, and the step length is set to 150. Since only one-third of the total shots are used to update the gradient in each iteration, the computational cost per iteration is reduced by approximately one-third compared to using all shots with the Adam optimizer. The inversion result obtained using the mini-batch Adam optimizer is shown in Fig.\ref{fig19}. Fig.\ref{fig20} shows the single-trace comparisons extracted from Fig.\ref{fig19}. As shown in Figs \ref{fig19}(d) and \ref{fig20}(d), the mini-batch Adam optimizer achieves the same level of accuracy in just 1500 iterations as the standard full-batch Adam optimizer does after 4500 iterations. The reduced number of iterations is attributed to the stochastic nature of mini-batch updates, which helps the optimization escape from local optima. Consequently, the mini-batch Adam optimizer proves to reduce the computational cost per iteration and significantly decreases the number of iterations required for accurate velocity estimation compared to the case without mini-batching.

\subsection{Marmousi2 model}
\begin{figure*}
\centering
\includegraphics[width=1\textwidth]{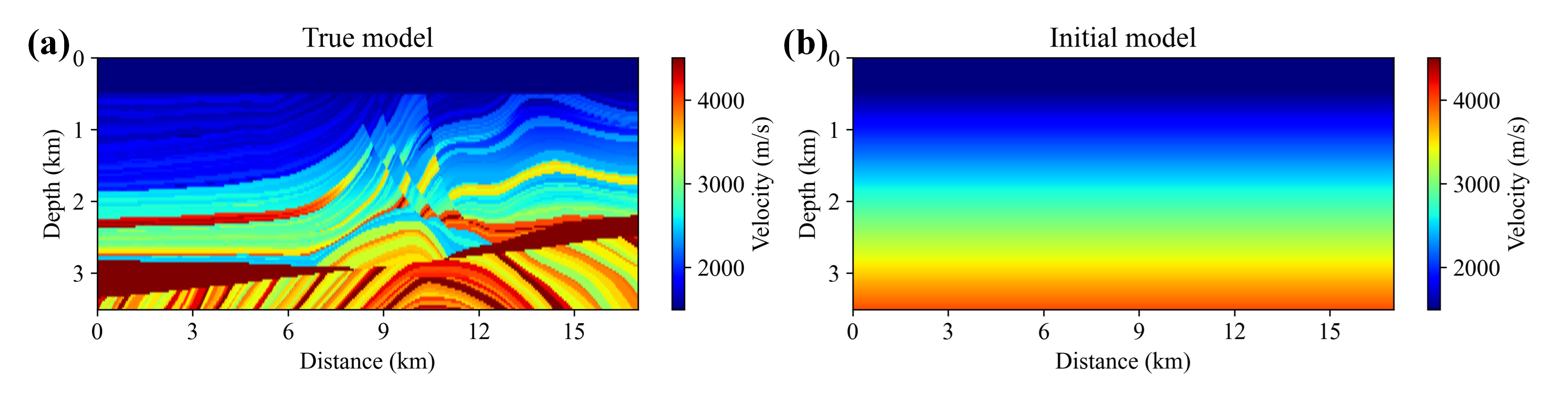}
\caption{(a) The true Marmousi2 velocity model, and (b) the corresponding initial model with a linearly increasing velocity.}
\label{fig21}
\end{figure*}

To further evaluate the robustness of the quasi-global optimization strategy, we apply FWI to the highly complex Marmousi2 model \cite{martin2006marmousi2}. Despite the geological complexity of this benchmark model, our results demonstrate that the GD optimizer can effectively overcome cycle skipping and recover an accurate velocity model, provided that a sufficient number of iterations are performed. 
The Marmousi2 model, shown in Fig. \ref{fig21}(a), consists of 567 × 117 grid points with a spatial interval of 30 m. A total of 32 shots are evenly distributed along the surface, with a shot interval of 540 m. For each shot, 567 receivers are uniformly placed along the surface with a receiver spacing of 30 m. A Ricker wavelet with a peak frequency of 5 Hz is used as the source. Each shot gather has a duration of 6 s with a temporal sampling interval of 3 ms. Fig. \ref{fig21}(b) shows the initial velocity model used for FWI, which is a gradient model with velocity increasing uniformly from top to bottom.

\begin{figure*}
\centering
\includegraphics[width=1\textwidth]{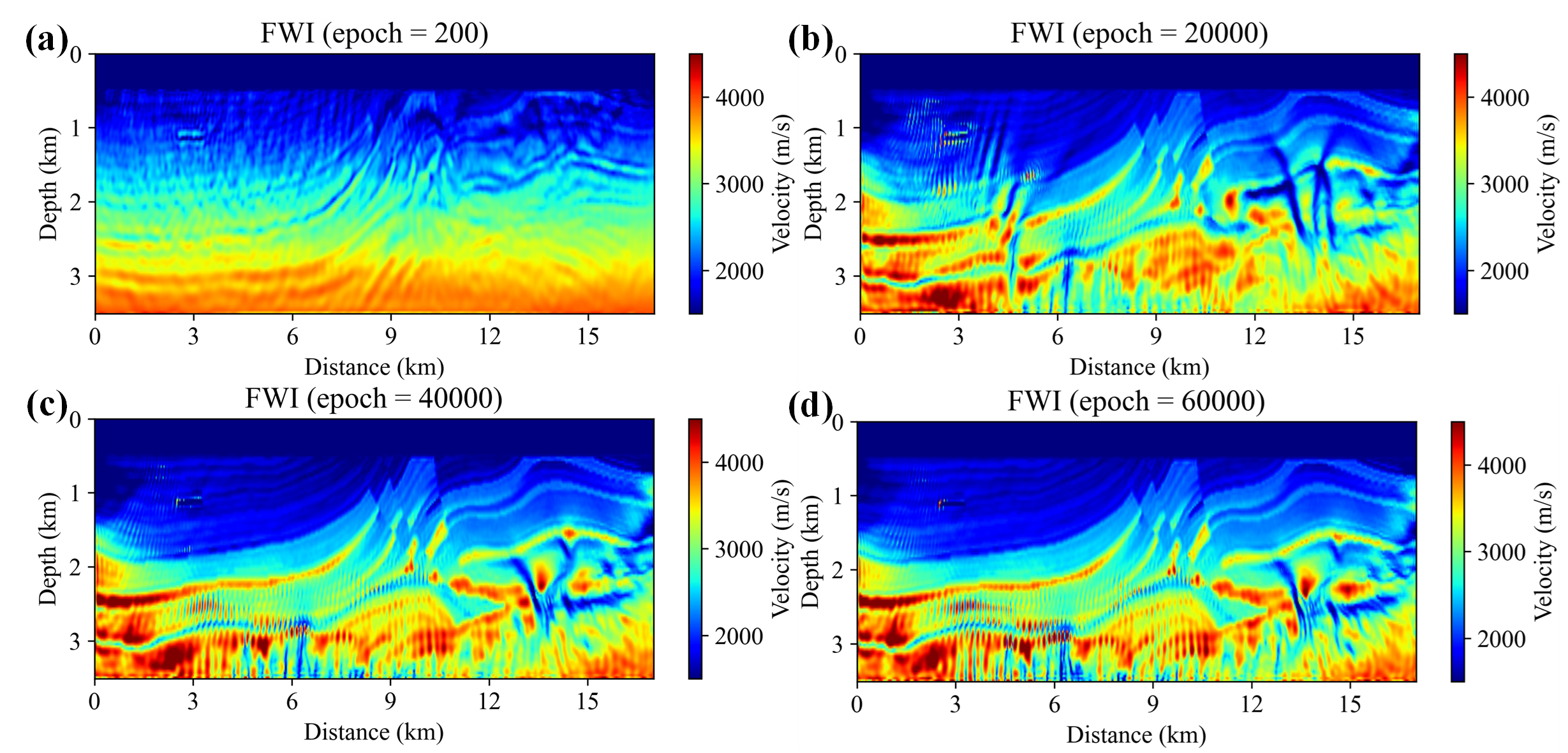}
\caption{FWI inversion results using the GD optimizer: (a)–(d) show the inverted results after 200, 20000, 40000, and 60000 iterations, respectively.}
\label{fig22}
\end{figure*}

\begin{figure*}
\centering
\includegraphics[width=1\textwidth]{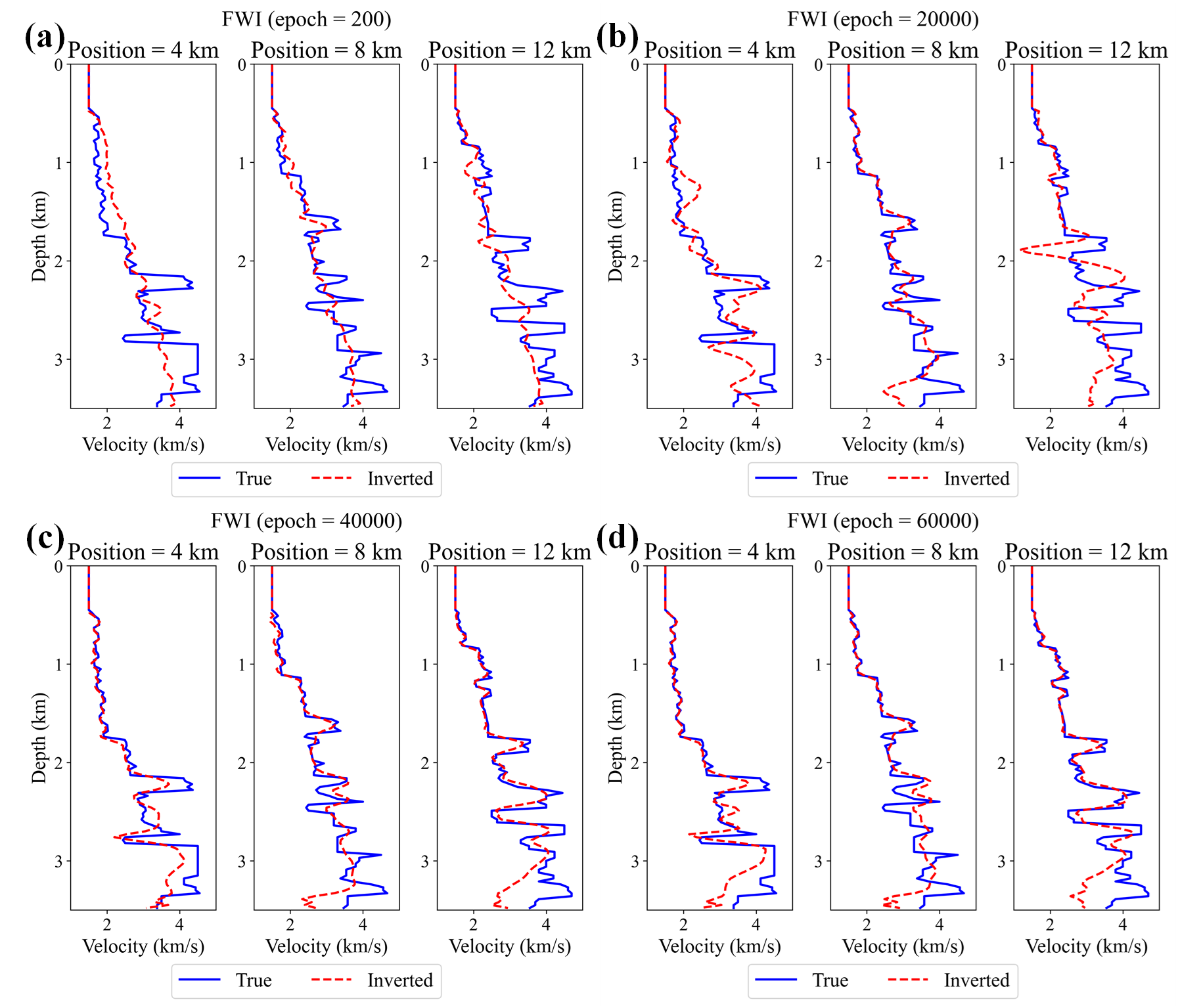}
\caption{One-dimensional velocity profiles extracted from horizontal locations 4 km, 8 km, and 12 km. Panels (a)–(d) display the profiles corresponding to Figs \ref{fig22}(a)–\ref{fig22}(d), respectively.}
\label{fig23}
\end{figure*}

\begin{figure*}
\centering
\includegraphics[width=0.7\textwidth]{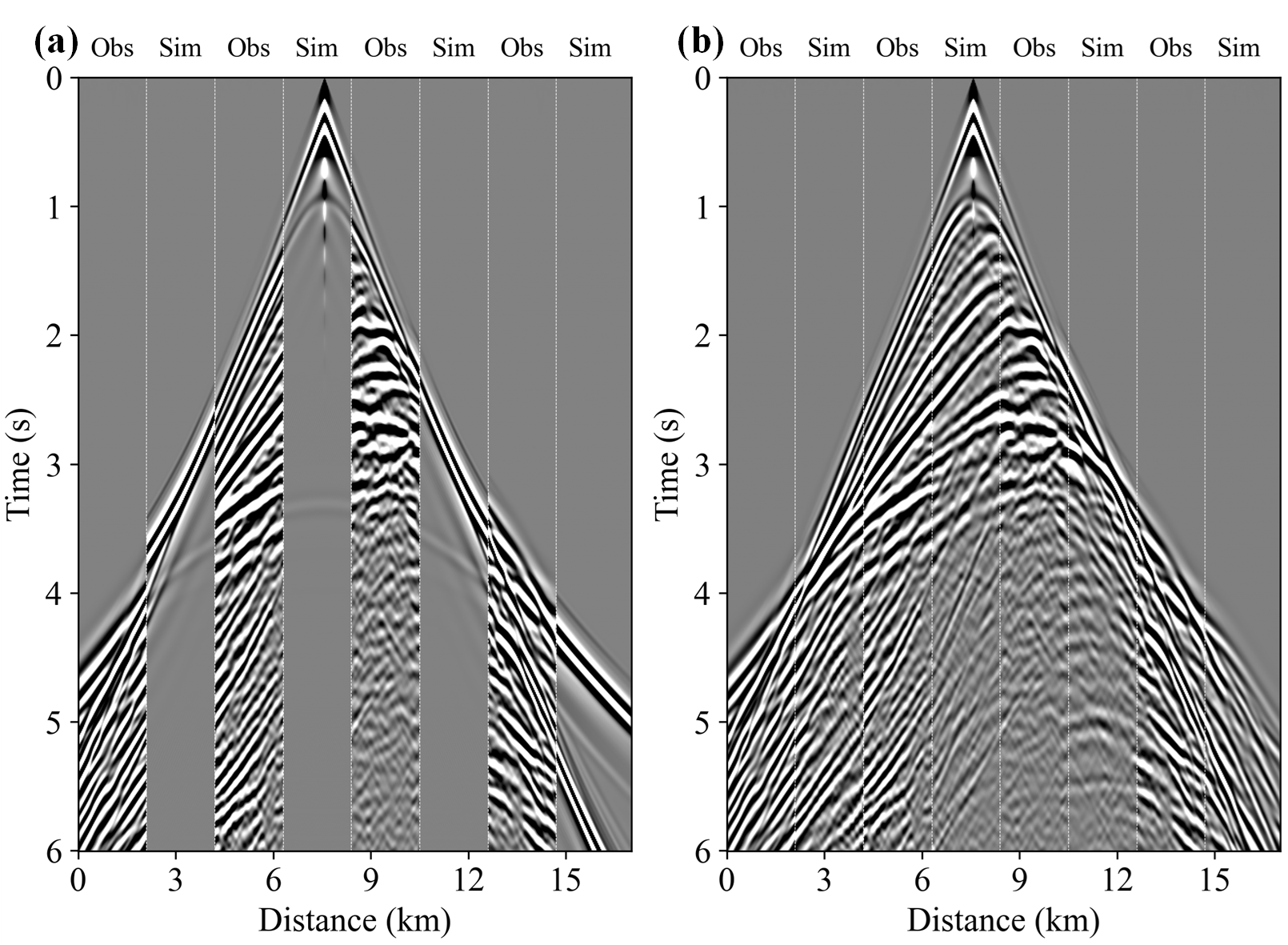}
\caption{Shot gather comparisons between the observed and simulated data for the 15th shot. Panels (a) and (b) show the simulated data computed using the inverted velocity models after 0 and 60,000 iterations, respectively. The panels labeled 'Obs' and 'Sim' represent the observed and simulated data, respectively.}
\label{fig24}
\end{figure*}

The inverted velocity models at iterations 200, 20000, 40000, and 60000 are shown in Fig. \ref{fig22}. Fig. \ref{fig23} displays a comparison of single-trace records extracted from the models in Fig. \ref{fig22}. In this case, FWI is performed using the GD optimizer and the step length is 30000. From the inversion results in Fig. \ref{fig22} and the trace comparisons in Fig. \ref{fig23}, it is evident that at iteration 200, there is a significant discrepancy between the inverted and true velocity models. However, as the number of iterations increases, the inversion accuracy improves progressively, reaching a high level of accuracy by iteration 60000. These results demonstrate that, even for complex model, FWI based on the GD optimizer can mitigate cycle skipping and yield high-accuracy inversion results, provided that a sufficient number of iterations are performed. Fig. \ref{fig24} shows a comparison between the observed and simulated data generated using the initial velocity model and the inverted model after 60000 iterations. There is evident cycle skipping between the observed data and the data simulated using the initial velocity model. In contrast, the data simulated with the velocity model obtained after 60000 iterations matches the observed data well, indicating that the inverted velocity model is very close to the true velocity model.

\begin{figure*}
\centering
\includegraphics[width=1\textwidth]{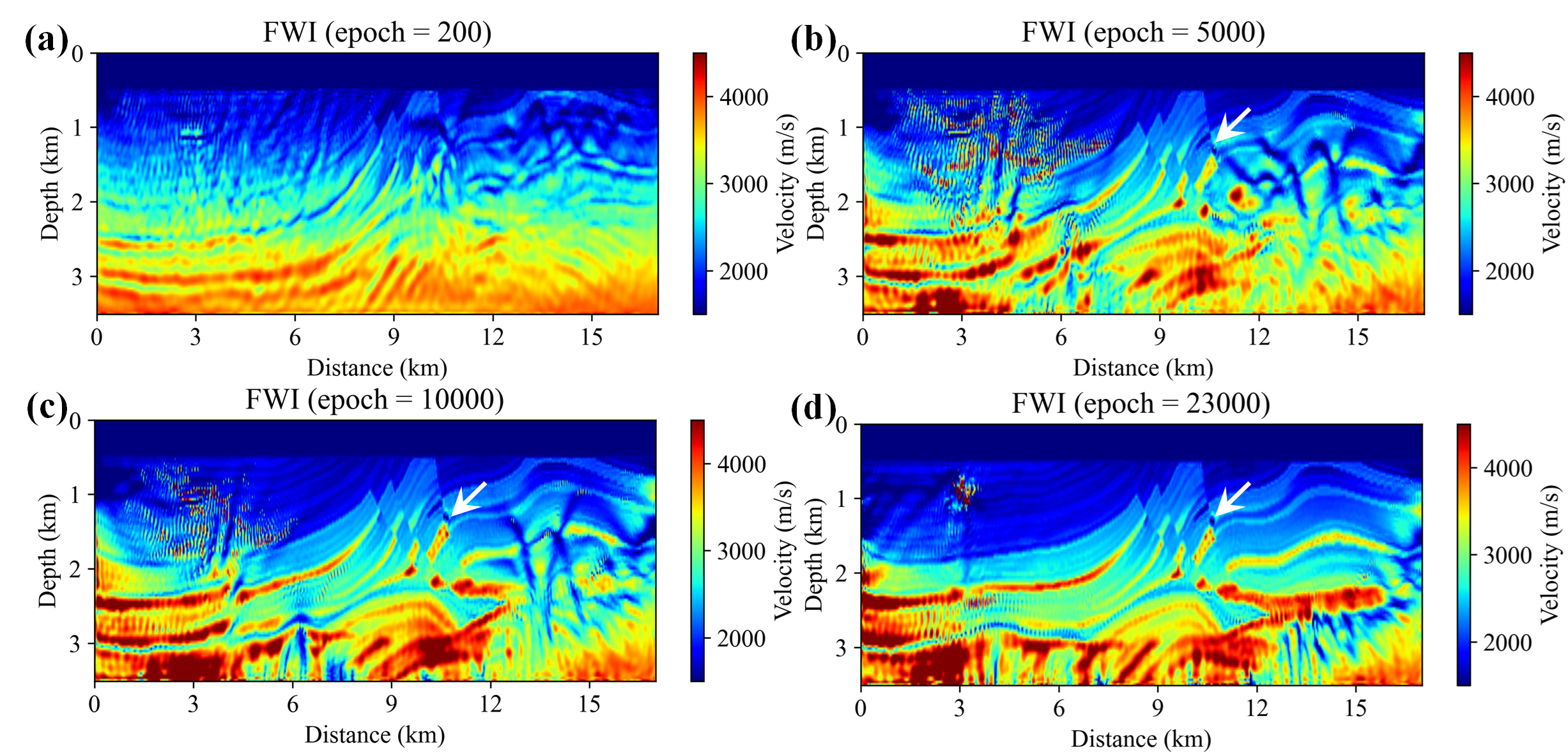}
\caption{FWI inversion results using the momentum-based GD optimizer: (a)–(d) show the inverted results after 200, 5000, 10000, and 23000 iterations, respectively. The white arrows indicate deep local minima.}
\label{fig25}
\end{figure*}

\begin{figure*}
\centering
\includegraphics[width=1\textwidth]{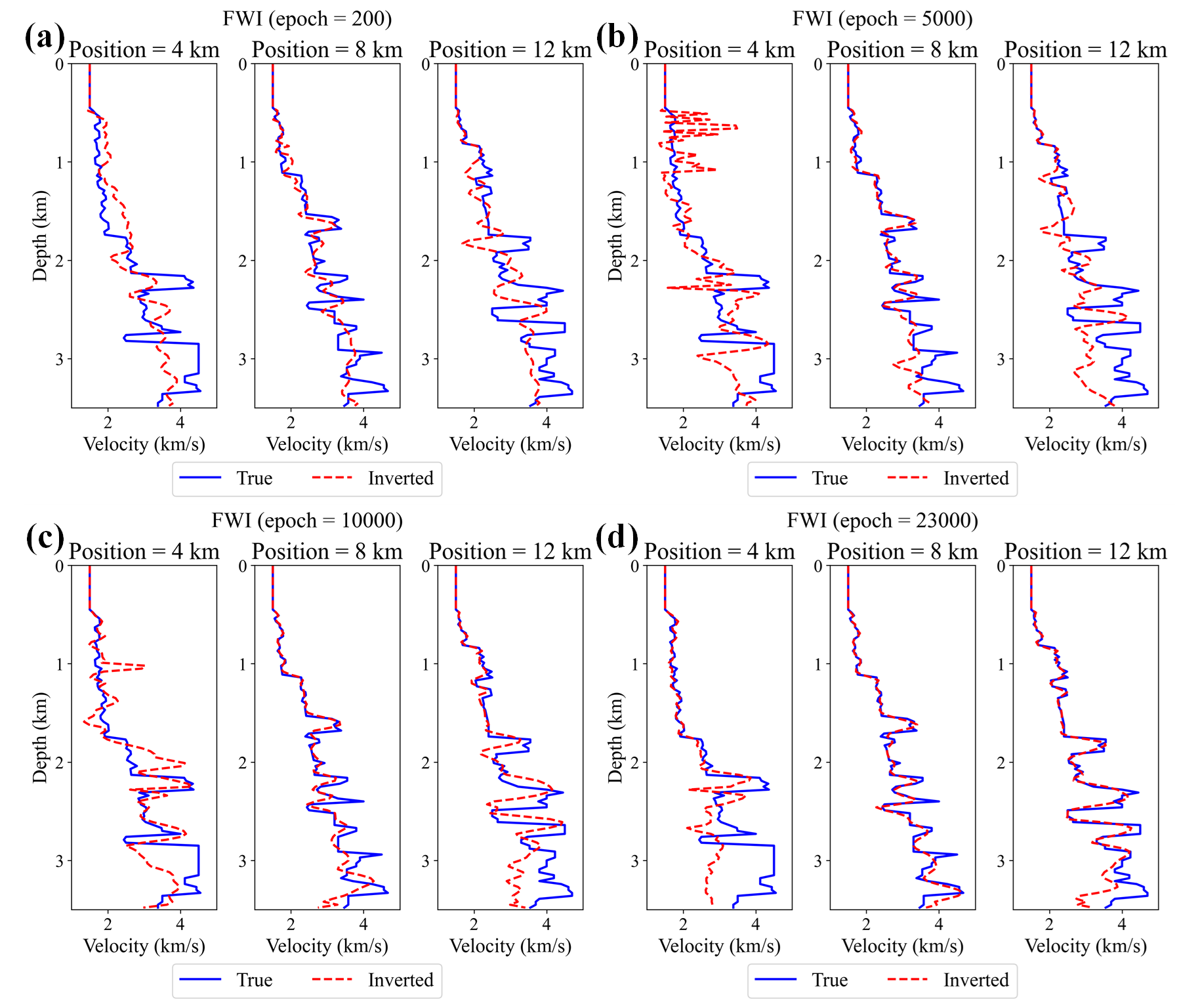}
\caption{One-dimensional velocity profiles extracted from horizontal locations 4 km, 8 km, and 12 km from Fig. \ref{fig25}. Panels (a)–(d) correspond to the profiles obtained from Figs \ref{fig25}(a)–\ref{fig25}(d), respectively.}
\label{fig26}
\end{figure*}

Fig. \ref{fig25} shows the inversion results obtained using the momentum-based GD optimizer, while Fig. \ref{fig26} shows single-trace comparisons extracted from Fig. \ref{fig25}. The step length is set to 10000. Compared with the result obtained using the standard GD optimizer (Figs \ref{fig22} and \ref{fig23}), the momentum-based GD optimizer accelerates convergence toward a more accurate inversion result. However, as indicated by the white arrows in Figs \ref{fig25}(b)-\ref{fig25}(d), there exists a local minimum that traps the optimization, preventing it from escaping even with additional iterations. We explain the underlying cause of this local minimum in the Discussion section.

\begin{figure*}
\centering
\includegraphics[width=1\textwidth]{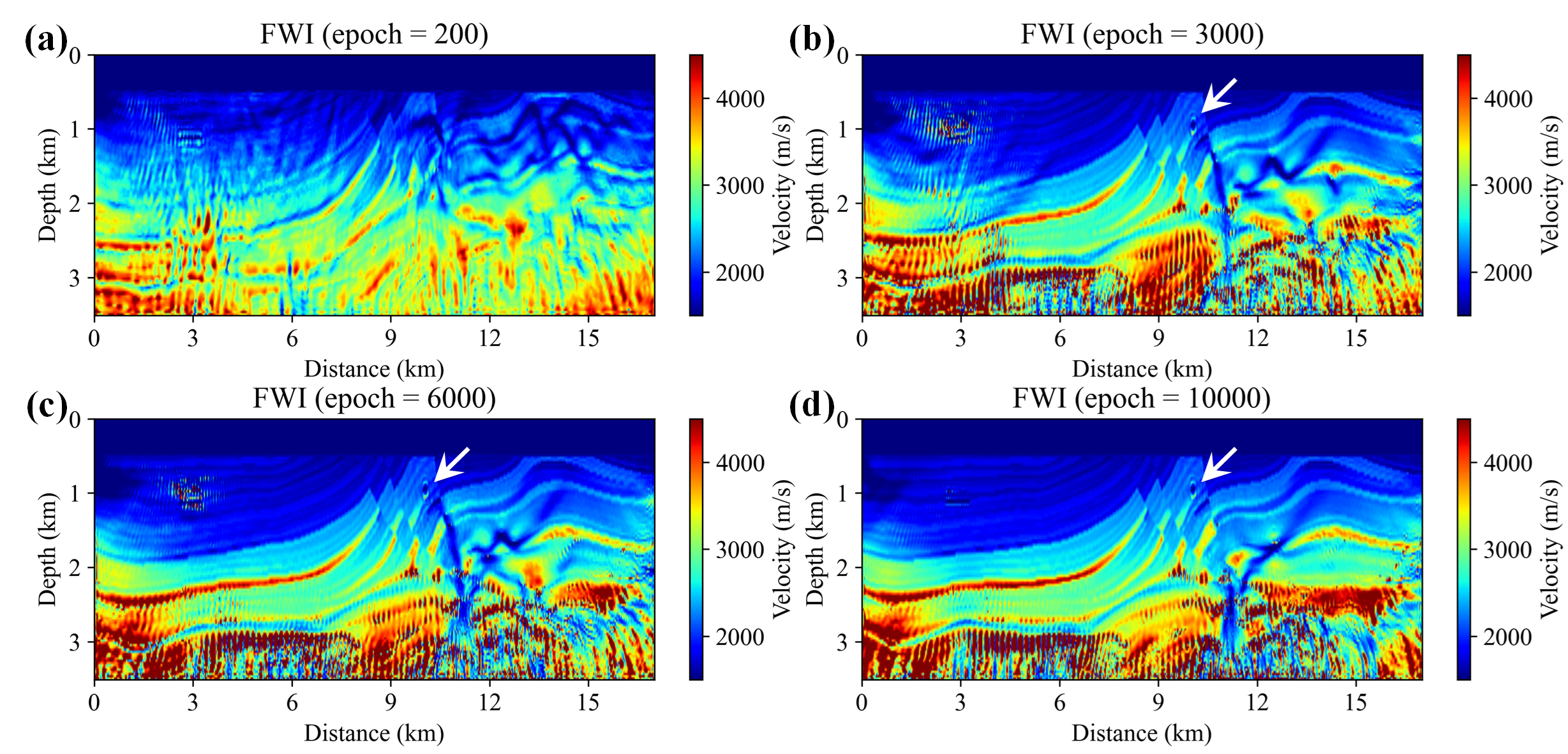}
\caption{FWI inversion results using the Adam optimizer: (a)–(d) show the inverted results after 200, 3000, 6000, and 10000 iterations, respectively. The white arrows indicate deep local minima.}
\label{fig27}
\end{figure*}

\begin{figure*}
\centering
\includegraphics[width=1\textwidth]{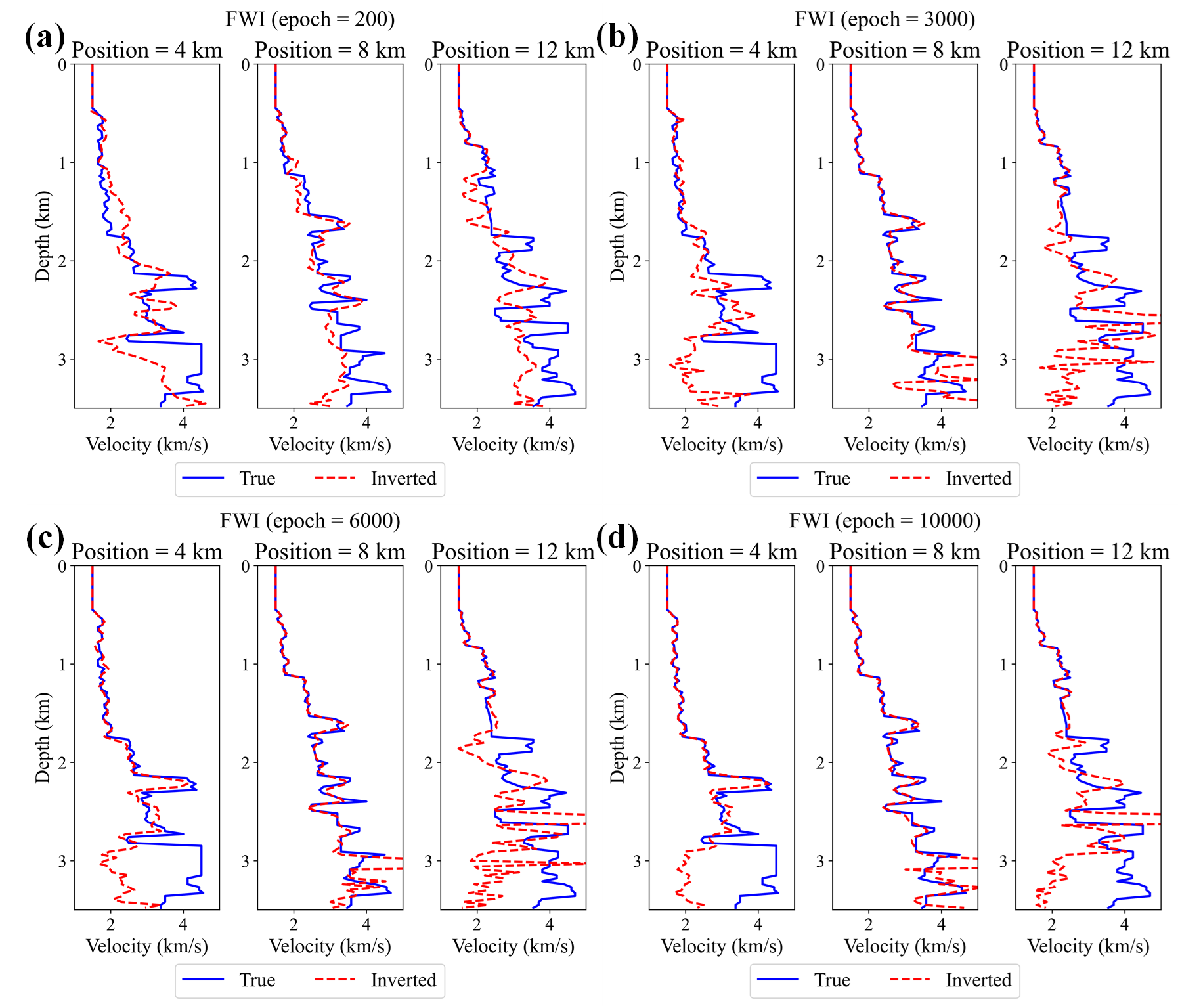}
\caption{One-dimensional velocity profiles extracted from horizontal locations 4 km, 8 km, and 12 km from Fig. \ref{fig27}. Panels (a)–(d) correspond to the profiles from Figs \ref{fig27}(a)–\ref{fig27}(d), respectively.}
\label{fig28}
\end{figure*}

The inversion results obtained using the Adam optimizer are shown in Fig. \ref{fig27}. Fig. \ref{fig28} shows single-trace comparisons extracted from Fig. \ref{fig27}. The step length is 60. By using the Adam optimizer, the number of iterations required to achieve an accurate result is further reduced compared to the inversion results obtained using the momentum-based gradient descent optimizer (Figs \ref{fig25} and \ref{fig26}). However, as indicated by the white arrow in Fig. \ref{fig27}, a local minimum still exists, preventing the inversion from accurately recovering the velocity in that area even with additional iterations. An explanation of the cause of this local minimum can be found in the Discussion section.

\begin{figure*}
\centering
\includegraphics[width=1\textwidth]{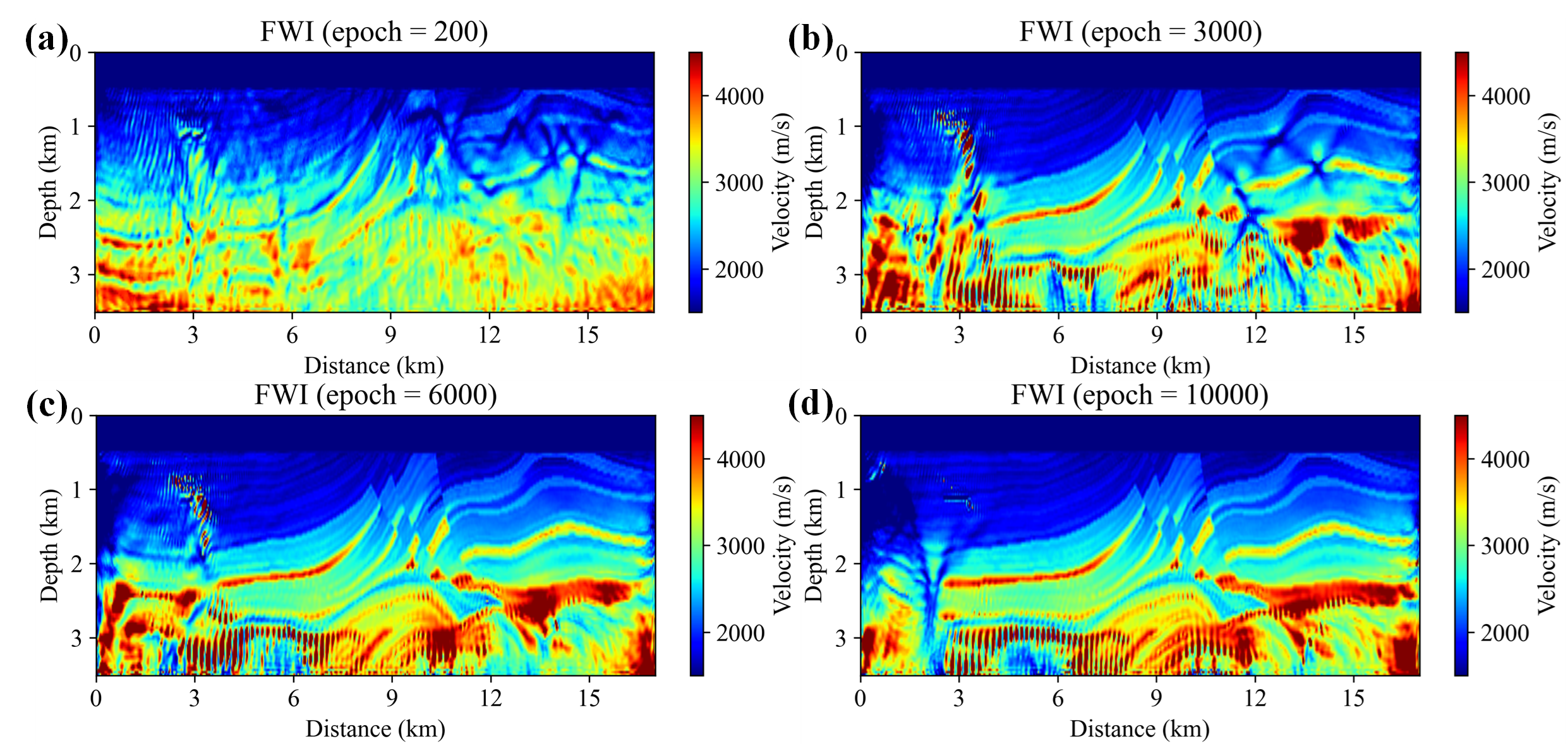}
\caption{FWI inversion results using the mini-batch Adam optimizer: (a)–(d) show the inverted results after 200, 3000, 6000, and 10000 iterations, respectively.}
\label{fig29}
\end{figure*}

\begin{figure*}
\centering
\includegraphics[width=1\textwidth]{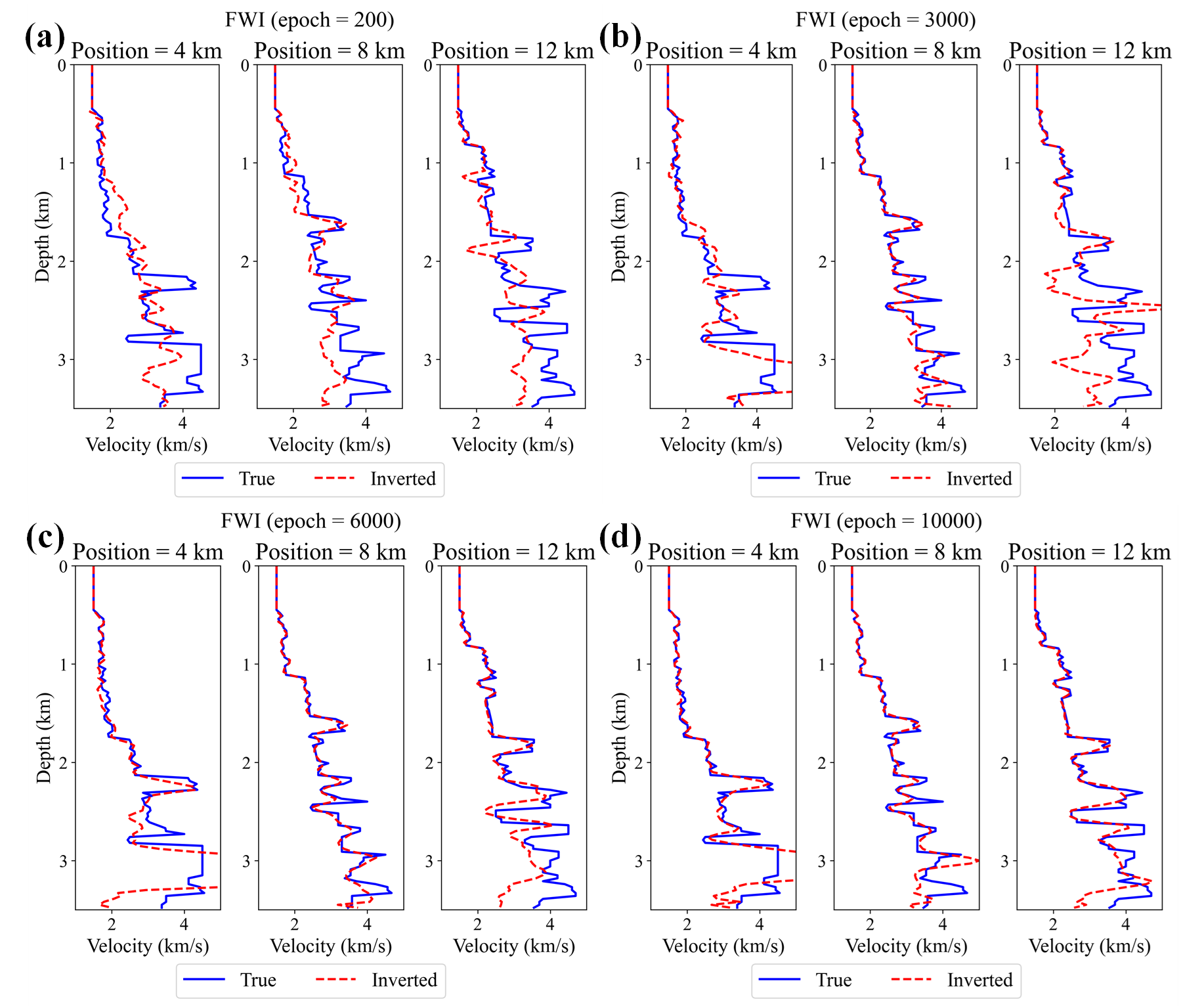}
\caption{One-dimensional velocity profiles extracted from horizontal locations 4 km, 8 km, and 12 km from Fig. \ref{fig29}. Panels (a)–(d) correspond to the profiles obtained from Figs \ref{fig29}(a)–\ref{fig29}(d), respectively.}
\label{fig30}
\end{figure*}

We also apply the mini-batch Adam optimizer for FWI, and the resulting inversion results are shown in Fig. \ref{fig29}. Fig. \ref{fig30} shows single-trace comparisons extracted from Fig. \ref{fig29}. The step length in this test is 60. This test demonstrates that the mini-batch Adam optimizer can significantly accelerate convergence toward an accurate velocity model in FWI compared to the full-batch Adam optimizer. A reasonably accurate inversion result is achieved within only 6000 iterations (Fig. \ref{fig29}(c)), although minor inaccuracies remain on the left side of the model. Meanwhile, the deeper local minima observed in Figs \ref{fig25} and \ref{fig27} (indicated by white arrows) are effectively eliminated.

\subsection{Field data}
To validate our insight that gradient descent optimizers can mitigate cycle skipping given a large number of iterations, we use a marine dataset acquired from northwestern Australia. The marine dataset comprises 116 shots, each generated by an airgun source and recorded over a duration of 7 seconds with a 1 ms sampling interval. The acquisition covers a lateral extent of roughly 20 km, with shots spaced approximately 90 meters apart. A towed streamer carrying 324 hydrophones, each spaced at 25 meters, was employed to record the seismic data. We resample the shot gathers to a 2 ms time interval. In this study, only data below 10 Hz is utilized. The model grid spacing is set to 25 meters. The source wavelet for each shot is inverted based on the known seawater velocity of 1500 m/s, utilizing the direct wave arrivals observed in the near-offset traces. 

\begin{figure*}
\centering
\includegraphics[width=1\textwidth]{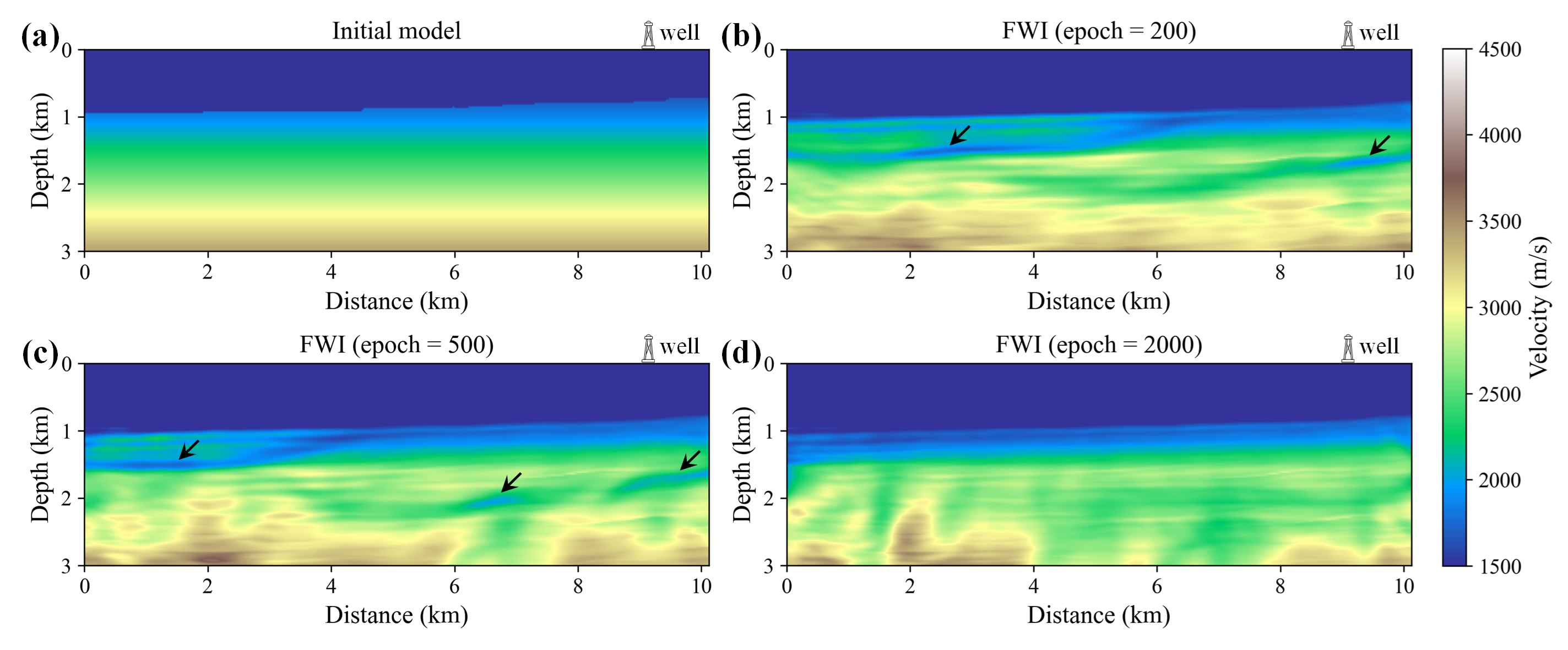}
\caption{Field data inversion: Initial model (a) and FWI inversion results after 200 (b), 500 (c), and 2000 (d) iterations.}
\label{figf1}
\end{figure*} 

\begin{figure*}
\centering
\includegraphics[width=0.36\textwidth]{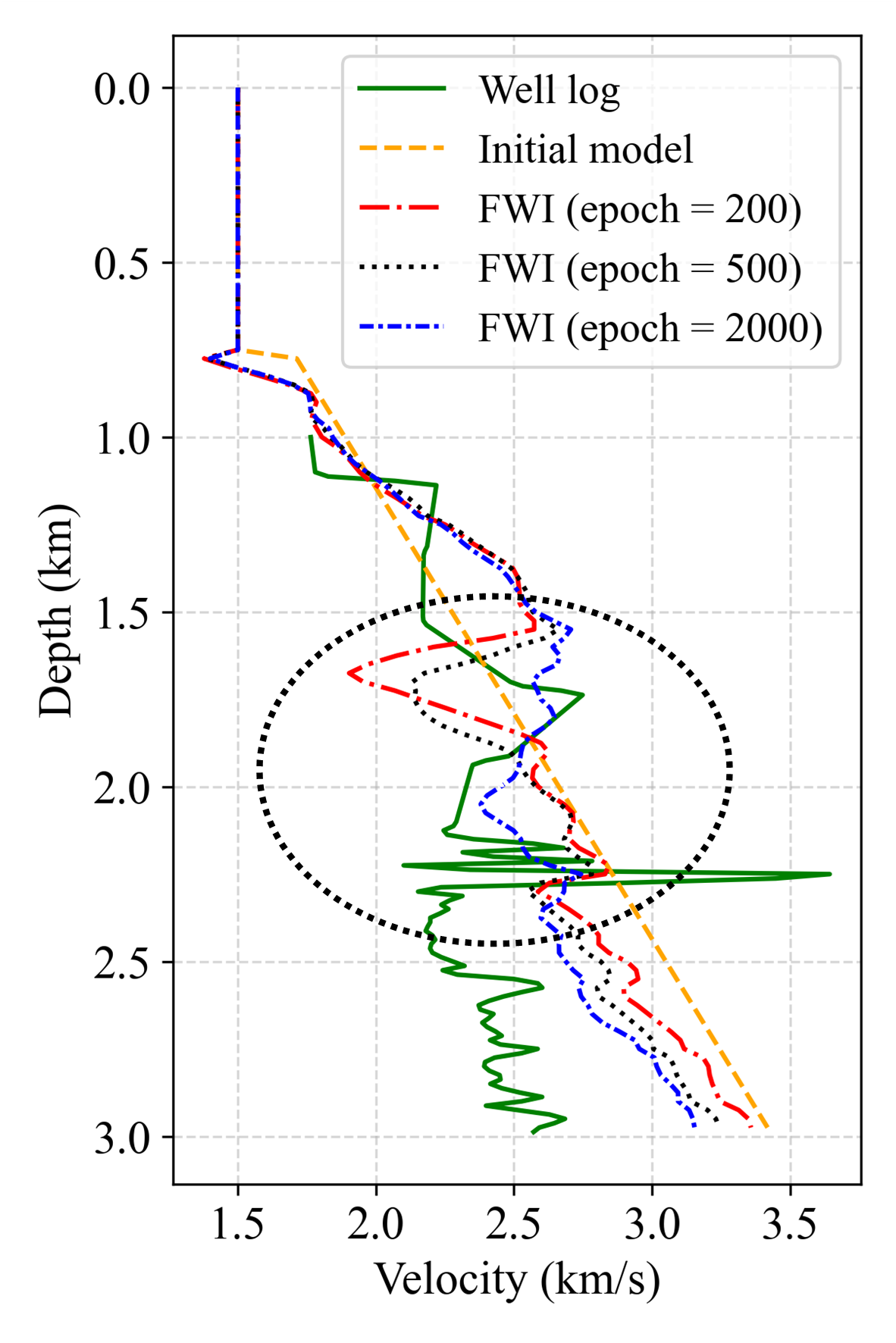}
\caption{Comparison of the velocity inversion results with the well log, where the well is positioned at 10.4 km.}
\label{figf2}
\end{figure*}

\begin{figure*}
\centering
\includegraphics[width=1\textwidth]{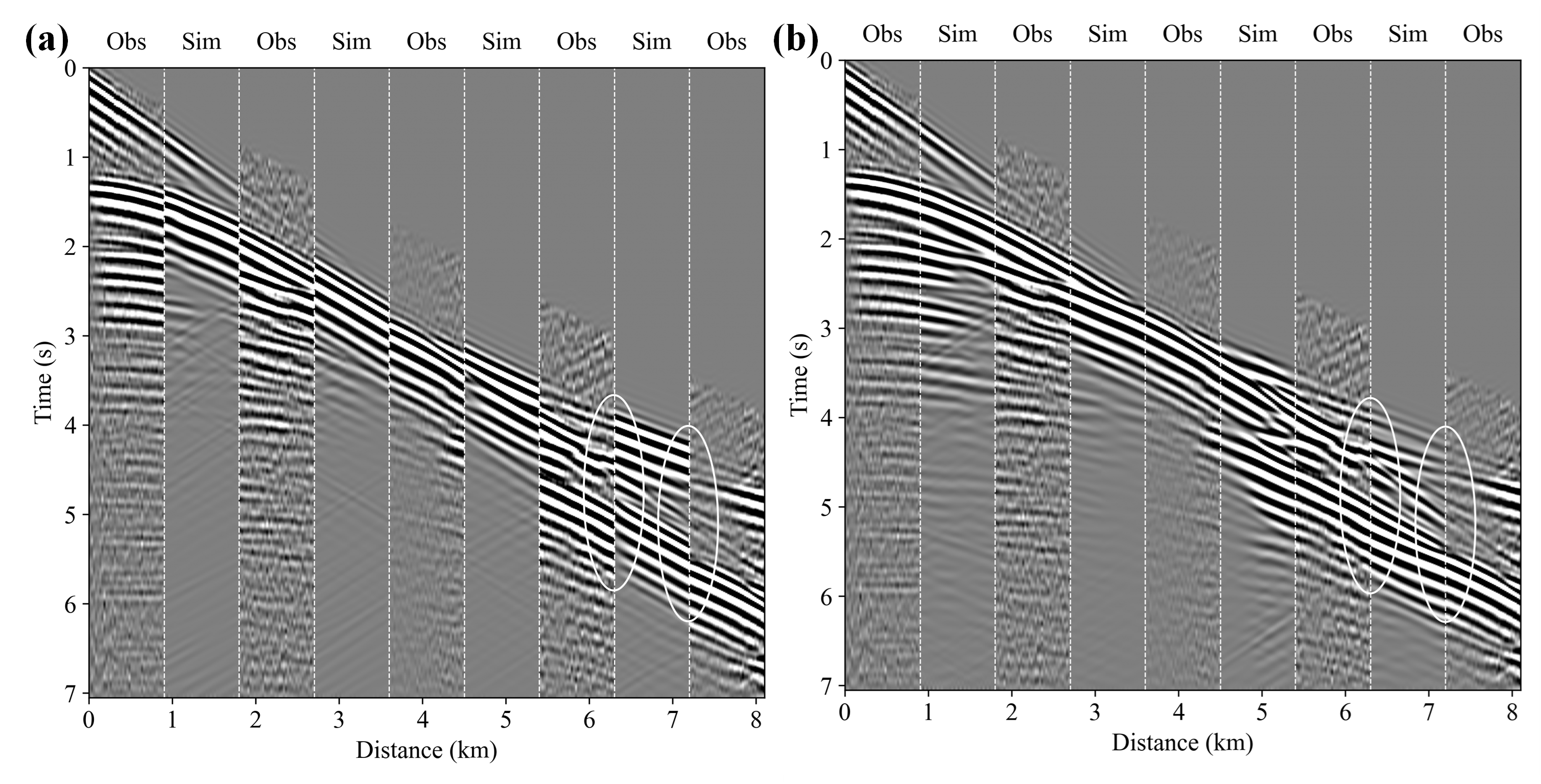}
\caption{Comparison of shot gathers between the observed data and the simulations computed using (a) the initial model and (b) the inversion model after 2000 iterations. In the figure, "Obs" and "Sim" denote the observed data and simulated data, respectively. The shot number is 50.}
\label{figf3}
\end{figure*}

As shown in Fig. \ref{figf1}(a), the initial model used for FWI is a linearly increasing model. This test employs a cross-correlation-based loss function \cite{choi2012application} and utilizes the Adam optimizer with a learning rate of 10. A mini-batch strategy is employed during the FWI inversion process, where 40 shots are randomly selected at each FWI iteration to compute the gradient. Figs \ref{figf1}(b)-\ref{figf1}(d) show the inversion results after 200, 500, and 2000 iterations, respectively. As highlighted by the black arrows in Figs \ref{figf1}(b) and \ref{figf1}(c), the inversion results exhibit the presence of local minima. However, as demonstrated in Fig. \ref{figf1}(d), these local minima are effectively eliminated after 2000 iterations. Fig. \ref{figf2} shows a comparison between the inversion results and the well log. As the number of iterations increases, the inverted velocity gradually converges toward the well log. After 2000 iterations, a good match is achieved, as indicated by the black dashed box in the Fig. \ref{figf2}. Fig. \ref{figf3} shows a comparison between the observed and simulated shot gathers. The simulated data generated using the initial velocity model exhibits cycle skipping relative to the observed data, particularly at far offsets (as highlighted by the white dashed box). In contrast, the simulated data based on the inversion result after 2000 iterations closely matches the observed data. This demonstrates that, even for field data, gradient descent optimizers with a large number of iterations can effectively mitigate cycle skipping.

% As shown in Fig. \ref{figf2}(d), the inversion results in the deeper parts of the model and near the left and right boundaries seem unreliable. This is likely due to poor illumination in these regions caused by the limited acquisition aperture. The synthetic tests presented above also confirm that, due to this limited aperture, the inversion results in the deeper and lateral regions of the model are less reliable. As demonstrated in the linearly increasing velocity model test, although more accurate velocities in these areas can be gradually achieved through additional iterations, obtaining satisfactory results would require a prohibitively large number of iterations. In addition, since only seismic data below 10 Hz are used in this inversion process, the reflected waves that carry information about the deeper velocity structures are filtered out. This is another reason why the inversion results in the deeper parts of the model are inaccurate. 

\clearpage

\section{DISCUSSION}

In this section, we first explain why gradient-based optimizers, combined with a large step length, can overcome cycle skipping after a sufficient number of iterations. We then discuss how to select an appropriate step length and analyze the robustness of this behavior in challenging scenarios, such as when frequencies below 5 Hz are missing. Next, we demonstrate that cycle skipping can be resolved using the L2 loss, provided that a sufficient number of iterations is used. Finally, we highlight a similar phenomenon in the field of machine learning, known as double descent.

\subsection{Explaining how gradient-based optimizers address cycle skipping and role of the step length}
\begin{figure*}
\centering
\includegraphics[width=1\textwidth]{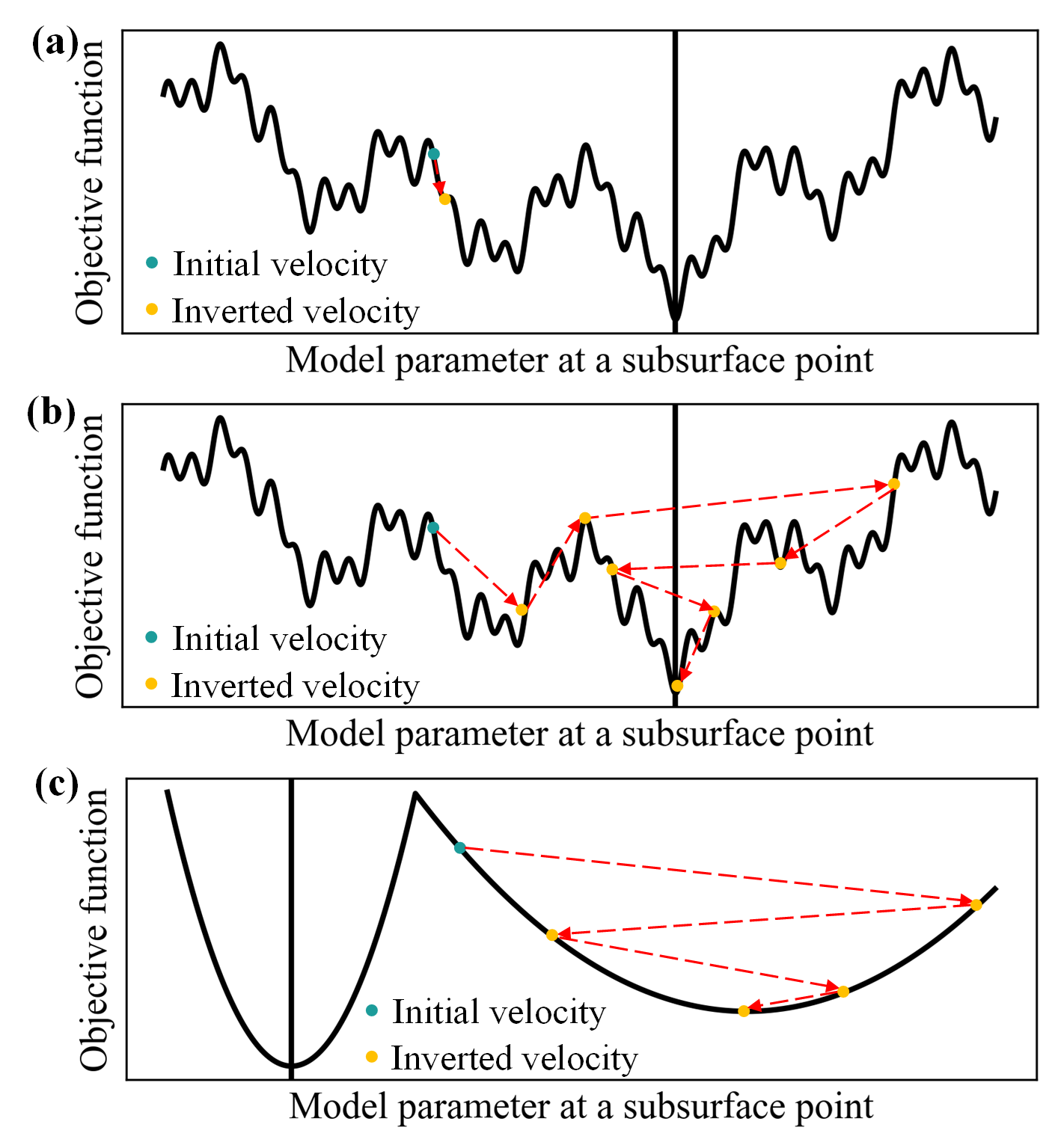}
\caption{Schematic diagrams of objective function landscapes at a subsurface grid point as a function of model parameters. Panel (a) corresponds to determining the step length via line search, while panels (b) and (c) correspond to determining the step length using a relative large fixed step length. The cyan and yellow dots indicate the initial and inverted velocities, respectively, while the red arrows show the direction of velocity updates.}
\label{fig32}
\end{figure*}

Local optimizers require the data misfit to decrease continuously, and therefore always use line search to determine the step length. As a result, when the inversion becomes trapped in a local minimum, the velocity model stops updating, as shown in Fig. \ref{fig32}(a). In contrast, here we allow the data misfit to oscillate rather than decrease monotonically. This enables the use of a relative large step length, which acts as a quasi-global optimization strategy to help escape local minima. As shown in Fig. \ref{fig32}(b), although the velocity inversion may encounter local minima, the large step length allows the inversion to escape them. Through multiple trial-and-error iterations, the velocity model ultimately converges to the global minimum. A big component of this is the nature of the loss function topology. While dimensions related to the deeper part is highly sensitive to values on the shallow attributable to our surface based recording, dimensions related to shallow parts of the model are not sensitive to values of velocity in the deep. Less sensitivity means more convexity and larger global minima along dimensions that are less sensitive to velocity changes on other areas. With a relatively sizable (but carefully picked) step length suitable for these shallow grid points dimension, the optimization is able to skip the smaller local minima and converge to a global one and the fit the data for such shallow grid points, which reduces the residuals. This feature slowly propagates downward as the converged shallow grid points become stationary, and the next depth experiences the same phenomenon. Interestingly, this fixed step length, considering the near similar wavelengths we are dealing with for the model, was appropriate. One may argue that the wavelengths (as velocity increase) increase with depth, which implies that an increase in the step length with iterations maybe be justified. As shown in Fig. \ref{fig32}(c), the local minimum at a subsurface point has a relatively wide basin. Even with a relatively large step length, the inversion may eventually converge to this local minimum, where the velocity update stagnates. This phenomenon can be observed in Figs. \ref{fig25} and \ref{fig27}, as indicated by the white arrows. However, when additional randomness is introduced into the inversion process—for example, by adopting a mini-batch strategy, as shown in Fig. \ref{fig29}—such local minima can be effectively mitigated.

The term “large step length” is relative: gradient-based local optimization methods typically require the data misfit to decrease monotonically, whereas here we allow the data misfit to oscillate. As a result, the step lengths used in this work are larger than those typically employed in local gradient-based optimization. However, the step length cannot be increased indefinitely. For the GD optimizer, the learning rate is generally chosen in the range of 10000–100000, while for the Adam optimizer, the step length is typically selected between 1 and 200.

\subsection{Performance in the absence of frequencies below 5 Hz}
\begin{figure*}
\centering
\includegraphics[width=1\textwidth]{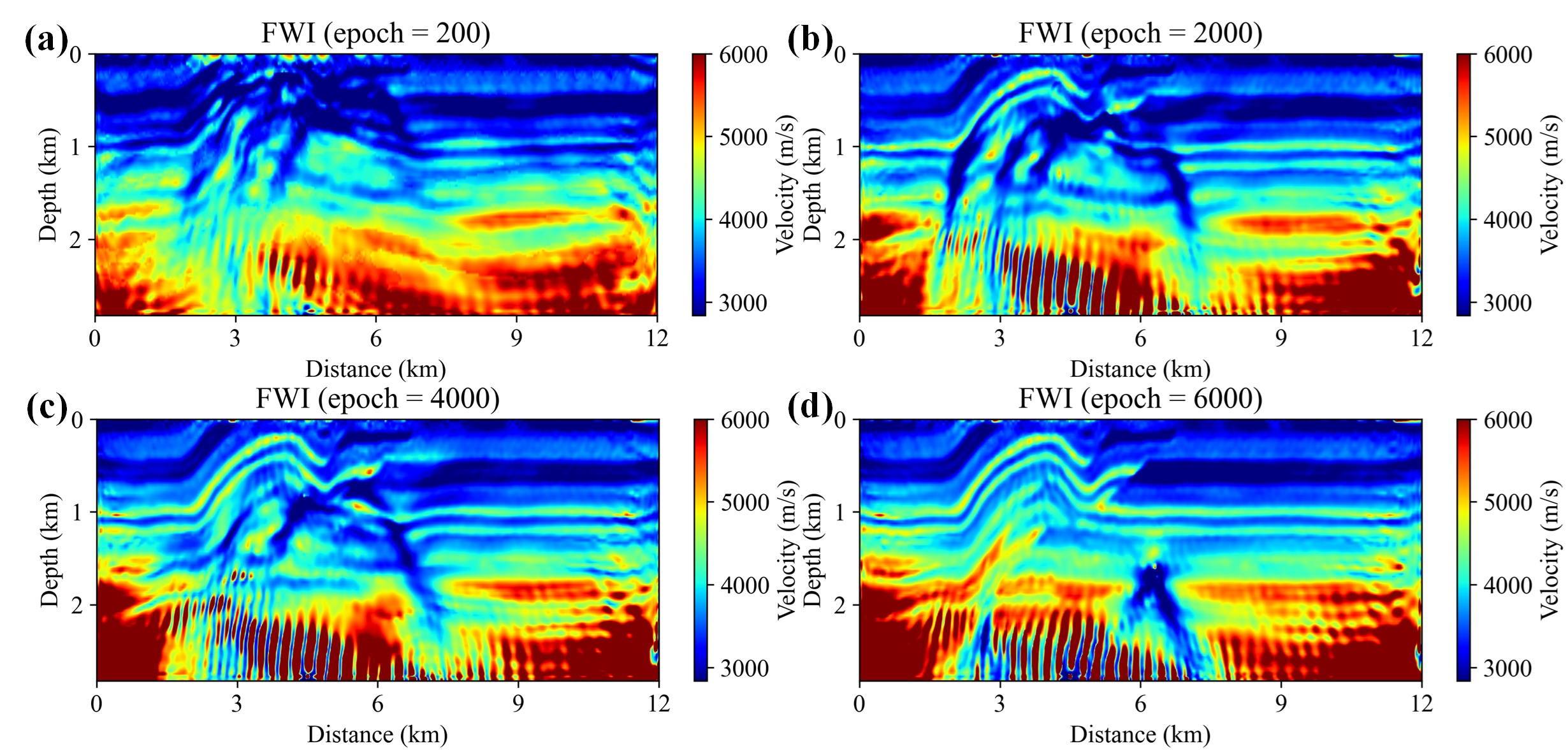}
\caption{FWI inversion results using the Adam optimizer: (a)–(d) are the inverted results after 200, 2000, 4000, and 6000 iterations, respectively. The observed data lacks frequency components below 5 Hz.}
\label{fig33}
\end{figure*}

\begin{figure*}
\centering
\includegraphics[width=1\textwidth]{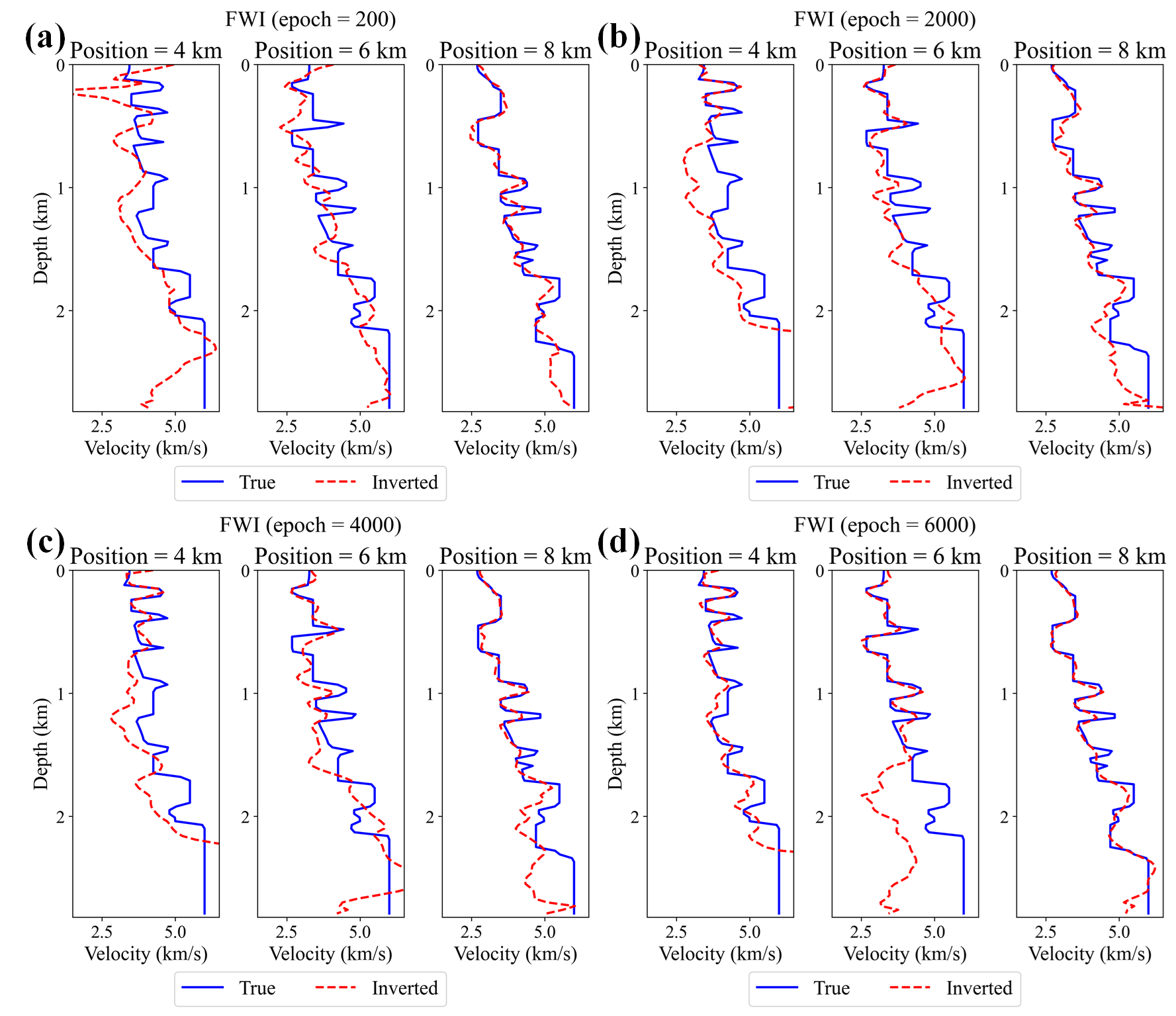}
\caption{One-dimensional velocity profiles extracted at horizontal distance of 4 km, 6 km, and 8 km from Fig. \ref{fig33}. Panels (a)–(d) correspond to the profiles obtained from Figs \ref{fig33}(a)–\ref{fig33}(d), respectively.}
\label{fig34}
\end{figure*}

We take the Overthrust model as an example to investigate whether the Adam optimizer can solve cycle skipping, given sufficient iterations, even in the absence of frequency components below 5 Hz. It is important to emphasize that the Adam optimizer is chosen here because it enables faster and more stable inversion toward the accurate velocity model, rather than implying that it is the only optimizer capable of addressing the cycle-skipping problem when the data lacks frequency components below 5 Hz. The model parameters and seismic acquisition configuration are the same as those used in Fig. \ref{fig17}, except that the seismic data here lack frequency components below 5 Hz. The step length used here is also set to 150. Fig. \ref{fig33} shows the FWI inversion results at different iterations, and Fig. \ref{fig34} shows the corresponding single-trace comparisons extracted from Fig. \ref{fig33}. From these figures, we observe that even though the observed data lack frequency components below 5 Hz, accurate inversion results can still be obtained using the Adam optimizer as the number of iterations increases. Although a low-velocity anomaly appears at a depth of 2 km in Fig. \ref{fig33}(d), this anomaly can be accurately resolved with more iterations.

% \subsection{Robustness to strong random noise}
% We further investigate whether the Adam optimizer can effectively mitigate cycle skipping in the presence of strong random noise in the observed data, provided that a sufficient number of iterations is allowed. The model, seismic acquisition configuration, and step length used here are the same as those in Fig. \ref{fig17}. The only difference is that we add random noise to the observed data, resulting in a signal-to-noise ratio (SNR) of 5 dB. Fig. \ref{fig35} shows the FWI inversion results at different iterations, and Fig. \ref{fig36} shows the corresponding 1D velocity profiles extracted from Fig. \ref{fig35}. As shown in these two figures, even with strong random noise in the observed data, the Adam optimizer can still yield accurate velocity inversion results, although it requires 3500 more iterations than in the noise-free case.

% \begin{figure*}
% \centering
% \includegraphics[width=1\textwidth]{Fig/fig35.png}
% \caption{FWI inversion results using the Adam optimizer: (a)–(d) are the inverted results after 200, 2000, 5000, and 8000 iterations, respectively. The SNR of the observed data is 5 dB.}
% \label{fig35}
% \end{figure*} 

% \begin{figure*}
% \centering
% \includegraphics[width=1\textwidth]{Fig/fig36.png}
% \caption{One-dimensional velocity profiles extracted at horizontal distance of 4 km, 6 km, and 8 km from Fig. \ref{fig35}. Panels (a)–(d) correspond to the profiles obtained from Figs \ref{fig35}(a)–\ref{fig35}(d), respectively.}
% \label{fig36}
% \end{figure*} 

\subsection{Can cycle skipping be solved using any loss function?}
\begin{figure*}
\centering
\includegraphics[width=1\textwidth]{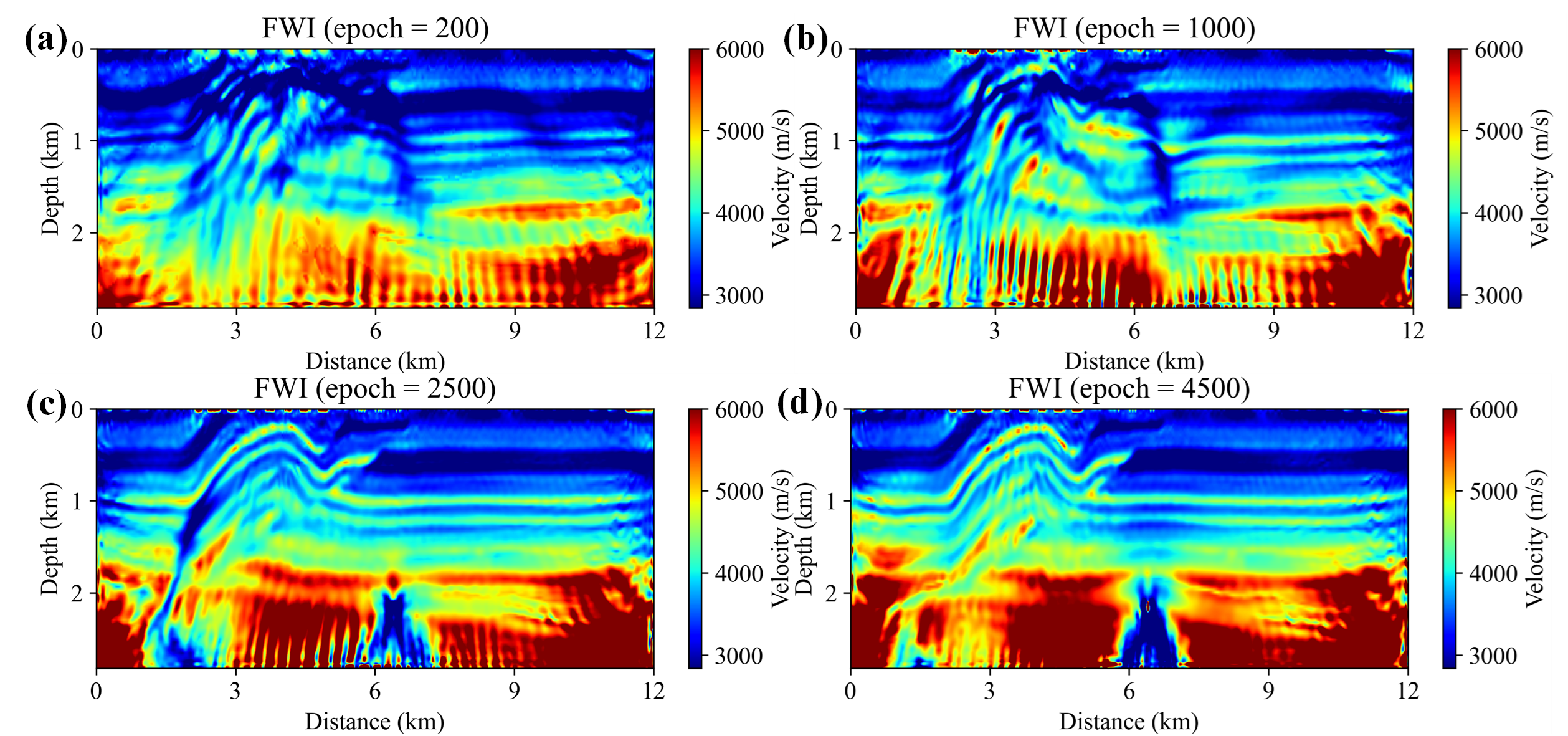}
\caption{FWI inversion results using the Adam optimizer: (a)–(d) are the inverted results after 200, 1000, 2500, and 4500 iterations, respectively. The L2 misfit function is used here.}
\label{fig37}
\end{figure*}

\begin{figure*}
\centering
\includegraphics[width=1\textwidth]{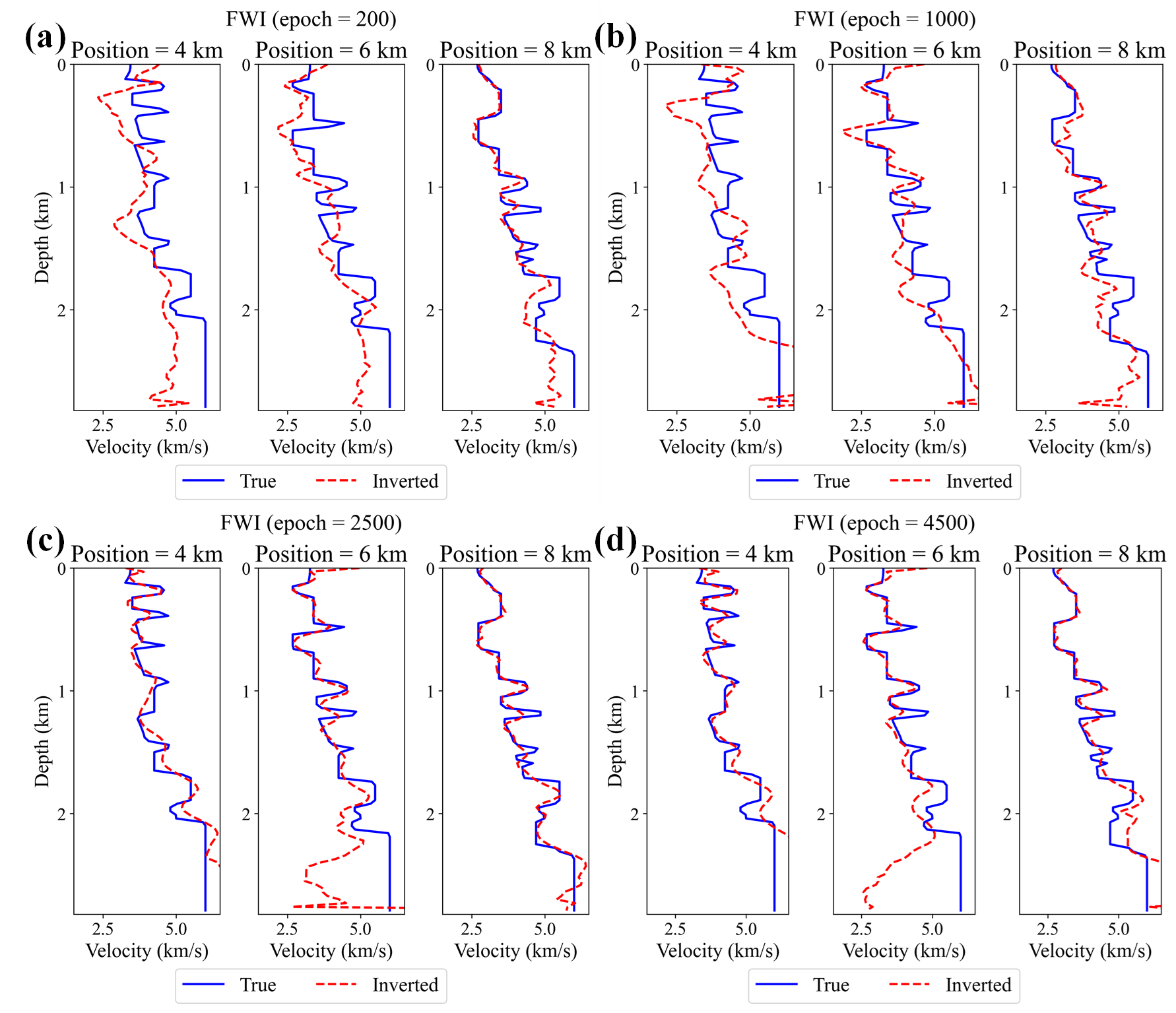}
\caption{One-dimensional velocity profiles extracted at horizontal distance of 4 km, 6 km, and 8 km from Fig. \ref{fig37}. Panels (a)–(d) correspond to the profiles obtained from Figs \ref{fig37}(a)–\ref{fig37}(d), respectively.}
\label{fig38}
\end{figure*}

To demonstrate that a large number of iterations can resolve cycle skipping in FWI regardless of the chosen objective function, we use the Overthrust model as an example. The model parameters, acquisition configuration, and step length are identical to those used in Fig. \ref{fig17}, except that the L2 misfit function (\(\Phi ({\bf{m}}) = \left\| {{{\bf{d}}_{syn}}({\bf{m}}) - {{\bf{d}}_{obs}}} \right\|_2^2,\)) is employed here. Seismic signals below 3 Hz are also removed. The inversion results are shown in Fig. \ref{fig37}, and 1D velocity profiles extracted from Fig. \ref{fig37} are shown in Fig. \ref{fig38}. These two figures demonstrate that, even with the L2 objective function, cycle skipping can be successfully mitigated, leading to accurate inversion results. Based on this test (using the L2 loss function) and our previous field data test (employing a cross-correlation loss function), we conclude that combining gradient-based optimizers, large learning rates, and a sufficient number of iterations is effective in addressing cycle skipping and is independent of the specific objective function used.

% \subsection{Insights gained from experimental results}
% Our numerical experiments suggest that, given a sufficient number of iterations, FWI is capable of mitigating cycle skipping and recovering an accurate velocity model. This implies that many of our previous efforts—such as using tomography to obtain a better initial model, adopting advanced objective functions, and implementing multi-scale strategies—can be interpreted as ways to accelerate convergence toward the accurate solution. When working with field data, if local minima still persist despite using a tomography-based initial velocity model, advanced objective functions, and multi-scale strategies, a practical and often effective solution is to simply increase the number of iterations. This extended optimization process can facilitate further model refinement and potentially guide the inversion toward a more accurate solution. We observe that as the number of iterations increases, the accuracy of the velocity model gradually improves. This may suggest that areas with frequent velocity updates during iterations are less reliable, while regions with minimal velocity variation are more trustworthy. This observation may help guide the development of new methods for uncertainty analysis in FWI.

\subsection{A similar phenomenon in machine learning: double descent}
\begin{figure*}
\centering
\includegraphics[width=1\textwidth]{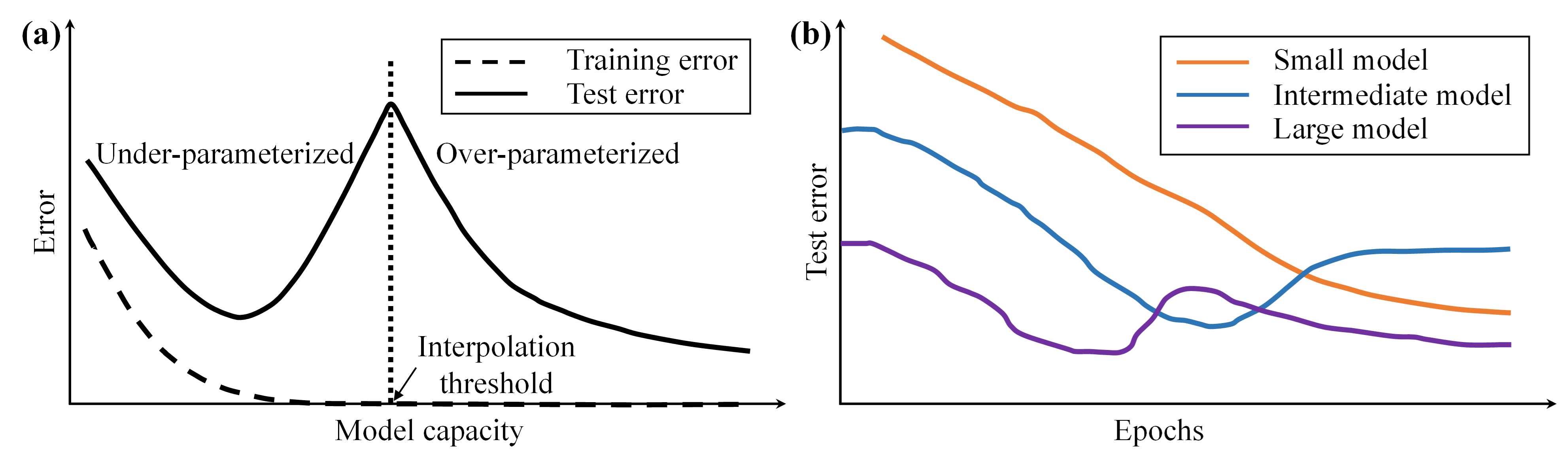}
\caption{Graphical depiction of the double descent curve: (a) training and test errors as functions of model capacity (adapted from \cite{cherkassky2024understand}), (b) test error as a function of the number of training iterations (Adapted from \cite{nakkiran2021deep}.)}
\label{fig41}
\end{figure*}

In the field of machine learning, a phenomenon known as double descent has been recently investigated \cite{belkin2019reconciling}, wherein increasing the number of training iterations can enhance a model’s generalization performance when the model is sufficiently over-parameterized (in our terms, not evenly illuminated, like shallow and deep). Double descent refers to a counterintuitive behavior of the test error curve that exhibits two distinct declines during training, thereby challenging the classical bias–variance trade-off in machine learning theory \cite{belkin2019reconciling}. Traditionally, as model capacity increases, training error monotonically decreases, while test error follows a U-shaped curve—initially decreasing due to improved expressivity, but eventually increasing due to over-fitting. However, the double descent phenomenon reveals that as model capacity continues to grow—particularly in the context of deep learning and highly over-parameterized neural networks—the test error may decrease again after reaching the interpolation threshold, resulting in a “double U-shaped” curve, as shown in Fig. \ref{fig41}(a). In our context, it implies that we first diverge and converge for the deeper part of the model. There are two common forms of double descent: (1) As the complexity of a model increases (e.g., through deeper architectures or a larger number of parameters), the test error initially decreases (under-fitting phase), then increases near the interpolation threshold (over-fitting phase), and finally decreases again in the over-parameterized regime, as depicted in Fig. \ref{fig41}(a); (2) Epoch-wise double descent: For a fixed over-parameterized model, increasing the number of training iterations may cause the test error to initially decrease, then rise due to over-fitting, and ultimately decrease again after further training, entering a regime of smoother and more stable fitting, as shown in Fig. \ref{fig41}(b). This phenomenon is of great importance because it suggests that larger models trained for longer periods do not necessarily suffer from over-fitting. On the contrary, over-parameterized models—when trained with appropriate optimization strategies—can achieve superior generalization performance. For FWI, the deep weaker illuminated parts of the model is experiencing its own double descent as it awaits the shallow part of the model to converge, and like double descent in machine learning, this phenomenon requires a lot of iterations.

\clearpage

\section{Conclusion}
We demonstrated that gradient-based optimizers with a large step length, when executed for a sufficient number of iterations, can effectively mitigate cycle skipping and ultimately produce high-accuracy inversion results. This conclusion is consistently supported across a range of models, including a linearly increasing model, the Overthrust model, the Marmousi2 model, and a marine field dataset. Importantly, even in challenging scenarios—such as the absence of low-frequency seismic data below 5 Hz—gradient-based optimizers are still able to produce reasonably accurate inversion results. Furthermore, our numerical experiments reveal that the basic gradient descent (GD) optimizer typically requires a substantial number of iterations to resolve cycle skipping. By contrast, incorporating momentum into the GD optimizer markedly reduces the number of iterations needed to achieve satisfactory results. Building on this, the Adaptive Moment Estimation (Adam) optimizer—which integrates both momentum and an adaptive learning rate—demonstrates even faster convergence toward accurate inversion. In addition, adopting a mini-batch strategy in conjunction with the Adam optimizer not only reduces the computational cost per iteration but also decreases the number of iterations required to obtain a reliable velocity model. Finally, we observe that the inversion accuracy naturally progresses from the shallow to the deeper parts as the number of iterations increases.

\section{Acknowledgments}
The authors sincerely appreciate the support from KAUST and the DeepWave Consortium sponsors and extend their thanks to the SWAG group for providing a collaborative research environment. The authors gratefully acknowledge the Supercomputing Laboratory at KAUST for providing the computational resources used in this work.

\bibliography{references.bib}
\bibliographystyle{unsrt}

\end{document}